\documentclass[11pt]{article}

	\usepackage{tikz}
	\usetikzlibrary{automata, arrows.meta, positioning}
	\usepackage{geometry}
 \usepackage{longtable}
 \usepackage{multirow}
 \usepackage{mathtools}
  \usepackage{tablefootnote}
  \usepackage{threeparttable}

    \makeatletter
    \renewcommand\section{\@startsection {section}{1}{\z@}%
                                       {-3.5ex \@plus -1ex \@minus -.2ex}%
                                       {2.3ex \@plus.2ex}%
                                       {\normalfont\fontfamily{phv}\fontsize{16}{19}\bfseries}}
    \renewcommand\subsection{\@startsection{subsection}{2}{\z@}%
                                         {-3.25ex\@plus -1ex \@minus -.2ex}%
                                         {1.5ex \@plus .2ex}%
                                         {\normalfont\fontfamily{phv}\fontsize{14}{17}\bfseries}}
    \renewcommand\subsubsection{\@startsection{subsubsection}{3}{\z@}%
                                        {-3.25ex\@plus -1ex \@minus -.2ex}%
                                         {1.5ex \@plus .2ex}%
                                         {\normalfont\normalsize\fontfamily{phv}\fontsize{14}{17}\selectfont}}
    \makeatother
	\usepackage{amsmath}
 \usepackage{titlesec}
 \usepackage{abstract}
\titleformat{\section}{\centering\large}{\thesection. }{0em}{\MakeUppercase}
 \titleformat{\subsection}{\large}{\thesubsection. }{0em}{\MakeUppercase}
 \renewcommand{\thesection}{\Roman{section}} 
\renewcommand{\thesubsection}{\thesection.\Alph{subsection}}
 \usepackage{mathptmx}
 \usepackage{setspace} 
\usepackage{subcaption}
 \usepackage{graphicx}

    \usepackage{esint}

    \usepackage{amssymb}
	\usepackage[english]{babel}
	\usepackage{graphicx}
	\usepackage{enumerate}
	\usepackage{lineno}	\usepackage[colorlinks=true, allcolors=blue]{hyperref}
    \modulolinenumbers[1]
	\usepackage{xcolor}
	\usepackage{float}
 \usepackage{lscape}
 
\usepackage{authblk}
\usepackage[backend=biber, style=apa, maxnames=99]{biblatex}
\usepackage{url} 

\newcommand*{\CC}{%
  \textsf{C\kern-1ex C}%
  
}
\newcommand*{\RR}{%
  \textsf{I\kern-.3ex R}%
}
\newcommand*{\ZZ}{%
  \textsf{Z\kern-1ex Z}%
}

\usepackage[capitalise,noabbrev,nameinlink]{cleveref} 

\title{Modelling Real-Life Cycling Decisions in Real Urban Settings Through Psychophysiology and LLM-Derived Contextual Data}

\author[1]{Maximiliano Rosadio Z.}
\author[1,3,4]{Angel Jimenez-Molina \thanks{ Corresponding author: Angel Jimenez-Molina; E-mail: angeljim@uchile.cl; Phone: +56 2 2978 0543; Address: Av. Beauchef 851 Of 711, Santiago, Chile, PC 8370458.}}
\author[2,4]{Bastián Henríquez}
\author[2]{Paulina Leiva}
\author[4,5]{Ricardo Hurtubia}
\author[2]{Ricardo De La Paz Guala}
\author[2]{Leandro Gayozo}
\author[2,4]{C. Angelo Guevara}

\affil[1]{Department of Industrial Engineering, Universidad de Chile.}
\affil[2]{Department of Civil Engineering, Universidad de Chile.}
\affil[3]{Data \& Artificial Intelligence Initiative - IDIA, Universidad de Chile.}
\affil[4]{Instituto Sistemas Complejos de Ingeniería, ISCI.}
\affil[5]{Department of Transport Engineering and Logistics \& School of Architecture, Pontificia Universidad Católica de Chile.}
\date{}

\begin{document}

\maketitle

\begin{abstract}
Measuring emotional states in transportation contexts is an emerging field. Methods based on self-reported emotions are limited by their low granularity and their susceptibility to memory bias. In contrast, methods based on physiological indicators provide continuous data, enabling researchers to measure changes in emotional states with high detail and accuracy. Not only are emotions important in the analysis, but understanding what triggers emotional changes is equally important. Uncontrolled variables such as traffic conditions, pedestrian interactions, and infrastructure remain a significant challenge, as they can have a great impact on emotional states. Explaining the reasons behind these emotional states requires gathering sufficient and proper contextual data, which can be extremely difficult in real-world environments. This paper addresses these challenges by applying an innovative approach, extracting contextual data (expert annotator level) from recorded multimedia using large language models (LLMs). In this paper, data are collected from an urban cycling case study of the City of Santiago, Chile. The applied models focus on understanding how different environments and traffic situations affect the emotional states and behaviors of the participants using physiological data. Sequences of images, extracted from the recorded videos, are processed by LLMs to obtain semantic descriptions of the environment. These discrete, although dense and detailed, contextual data is integrated into a hybrid model, where fatigue and arousal serve as latent variables influencing observed cycling behaviors (inferred from GPS data) like waiting, accelerating, braking, etc. The study confirms that cycling decisions are influenced by stress-related emotions and highlights the strong impact of urban characteristics and traffic conditions on cyclist behavior.\\

\vspace{7pt}


\end{abstract}

\newpage
\section{Introduction}\label{sec:intro}

In the past decade, rapid urbanization and population growth in metropolitan areas have significantly increased the demand for efficient and sustainable transportation systems. Among the alternatives, cycling stands out as a promising solution due to its efficiency, affordability, and environmental benefits (\cite{pucher2017cycling,oviedo2022arguments}). However, the widespread adoption of cycling as a primary mode of transport faces challenges, particularly related to perceived safety, comfort, and physical effort required by both current and potential cyclists (\cite{Gutirrez2020,gutierrez2025willingness}). Addressing these concerns requires a deeper understanding of cyclists’ psychological and emotional experiences in real-world urban environments. Traditionally, psychological factors in transportation studies have been evaluated using self-reported indicators, such as surveys or interviews. These methods have been effective in assessing key psychological states such as risk perception, modal predispositions, and lifestyle preferences (\cite{Hurtubia2014, munoz2016increasing}). However, such approaches lack granularity and are prone to response biases, including memory recall and subjective interpretations. Thus, these methods do not provide a detailed and granular understanding of how cyclists experience and react to their surroundings, and provide a limited understanding on how urban factors influence cycling decisions.

Recent advances in wearable biosensor technologies offer novel approaches. These devices enable the non-invasive and real-time capture of physiological data, such as heart rate (HR), heart rate variability (HRV), electrodermal activity (EDA), skin temperature (SKT), among others, providing objective insights into emotional states such as stress and comfort without relying on self-reports, as they do not suffer from response bias. Previous studies, such as \textcite{barriaRelatingEmotionsPsychophysiological2023,henriquez2025identifying}, have integrated physiological indicators into choice models, particularly in public transportation, to evaluate emotional and physiological responses of users in complex mobility scenarios. This shift toward physiological monitoring addresses key limitations of traditional methods and allows for a more nuanced understanding of user experiences. 

On the other hand, \textcite{bogaczModellingRiskPerception2021a} and \textcite{paschalidis2018modelling} estimated the effect that risk perception and stress has on the decisions of cyclists and car drivers. \textcite{bogaczModellingRiskPerception2021a} used electroelencephalography in a \textit{virtual reality} (\textit{VR}) cycling experiment and showed that factors such as distance to the intersection triggered risk perception and increased the probability of performing brake maneuvers. \textcite{paschalidis2018modelling} used \textit{electrodermal activity} (\textit{EDA}) in driving simulation scenarios, showing that the probability of gap acceptance increased under stress conditions due to time pressure. \textcite{millar2021space} collected data from a sample of 12 cyclists in the Netherlands, traveling on a cycle highway between two municipalities. They link the characteristics of the cycling environment (captured through cameras) with each individual's \textit{global position system} (\textit{GPS}) location, HR, SKT, and EDA, to understand the effect of environment visual stimuli on cyclists' arousal, monitoring speed, direction, distance to roads, and directional change. They show that roads with views of larger natural, recreational, agricultural, and forested areas have the effect of increasing the arousal of participants. However, this study does not provide evidence on the relationship between cyclist emotions and their on-route decisions. 

The referenced literature indicates that there is laboratory evidence suggesting a relationship between cyclists' emotions and their behavior, and field evidence of a relationship between the environment characteristics and the emotions of cyclists. However, there is no field evidence shedding light on the relationship between these three factors: individuals' emotions, the urban environment characteristics, and cyclists decisions. This article aims to fill this gap in the literature. 

Capturing all stimuli from the urban environment that could have an effect on the emotions of individuals is a challenge by itself. A feasible approach is to extract \textit{points of interest} (\textit{POI}) and infrastructure data from a map service (e.g. \textit{Google Maps}, \textit{Openstreetmap}, or \textit{Mapillary}), as done by \textcite{Klinkhardt2021} to estimate transport demand models. However, this approach does not capture real-time stimuli that may be present during the person's experience, such as congestion, proximity to other modes, weather, or pedestrian activity. To capture all these factors, it is necessary to implement ad hoc cameras for posterior image analysis. However, systematic video analysis challenges can make this prohibitive (e.g. \cite{millar2021space}). In this article, we show that this can be worked out using \textit{Large Language Model}s (\textit{LLM}s), since they are flexible enough to identify objects, situations, and interpret the context without additional training or manual data labeling requirements.

This article aims to contribute by integrating the use of LLMs to capture detailed contextual information extracted from video recording, in a discrete choice model that links urban features with latent psychological states and cyclists' decisions, employing psychophysiological data as indicators of psychological states. Data were collected from participants cycling on a fixed route in Santiago, Chile, while wearing an \textit{Empatica E4} device (\cite{garbarino2014empatica,Schuurmans2020}) and a \textit{GoPro} camera. This setup includes the Contextino kit, a tailored sensor platform to integrate measurements of air temperature, relative humidity, ambient noise level, CO2 concentration, and atmospheric pressure. An \textit{integrated choice and latent variable}  model (\textit{ICLV}, \cite{walkerGeneralizedRandomUtility2002}) was built, integrating latent arousal and fatigue as explanatory variables of cycling decisions (acceleration, deceleration, braking, maintaining speed and waiting). This novel methodology allows us to answer relevant behavioral questions regarding: P1. What urban environment and traffic characteristics affect cyclist stress and fatigue? P2. Are cycling decisions affected by latent psychological states of stress and fatigue? Despite previous laboratory evidence, this is the first study, to the extent of our knowledge, that provides field evidence to answer these questions. 

The article is structured as follows. After this introduction, \Cref{sec:background} presents a literature review, summarizing a background on \textit{discrete choice models} (\textit{DCM}), and the main challenges regarding the integration of DCM and \textit{artificial intelligence} (\textit{AI}) models. \Cref{sec:methods} describes the experiment design, instruments, data processing, and modeling approach. Then \Cref{sec:results} shows the main results and finally \Cref{sec:conclusion} draws the main conclusions, discusses the limitations of this study, highlights policy implications, and enunciates future research lines.


\section{Transport Choice Models: Background and Challenges in AI integration}\label{sec:background}

Transport behavior modeling is mainly approached from \textit{Random Utility Maximization} (\textit{RUM}, McFadden, 1974) theory. RUM models are used to understand and anticipate individuals' behavior in transport systems. They can be used to predict the transport mode a person will choose (car, bus, subway, bicycle, etc.) based on factors such as cost, travel time, and comfort (\cite{Train2002}), or used for traffic behavior modeling to represent drivers' acceleration, braking, or lane changing decisions (\cite{ahmed1999modeling,toledo2009estimation}).

RUM theory assumes that individuals are most likely to choose the alternative that maximizes their perceived utility. A basic RUM formulation considers that the utility or satisfaction an individual $n$ associates with an alternative $i$ can be represented by the utility function $U_{ni}=V_{ni}+\varepsilon_{ni}$ composed of two parts. The first is an observable part, known as \textit{systematic utility function}, is expressed linearly as $V_{ni}=\sum_k x_{nik}\beta_k$, where $x_{nik}$ is the value of an attribute $k$ of the alternative $i$, $\beta_k$ is the taste parameter of attribute $k$. The second part, $\varepsilon_{ni}$, is a \textit{random error term} that captures “idiosyncratic” variability. Then, the individual chooses the alternative with maximum utility, following:
\begin{equation}
	\label{ec1}
y_{ni}=1[U_{ni}=max_{j\in C}\{U_{nj}\} ]
\end{equation}
where $y_{ni}$ is an indicator function which equals 1 if individual $n$ chooses alternative $i$ and 0 otherwise, and $C$ represents the set of available alternatives. Here, the attributes are observable and are assumed to be known by the individuals as they have complete and perfect information.  

\cite{walkerGeneralizedRandomUtility2002} proposed an extension of the above explained RUM model to capture non-observable attributes: the ICLV model. Here, given an individual $n$ and an alternative $i$, the systematic utility function is given by $V_{ni}=\sum_{k}{x_{nik}\beta_k}+\sum_{l}{x_{nil}\beta_l}$, where $x_{nil}$ represents the value of a \textit{latent} (non-observable) value $l$ for individual $n$ and alternative $i$ (e.g. quality, or perceived environmental impact of an alternative), and $\beta_l$ is the taste parameter of latent value $l$. Although non-observable, latent variables can be measured within a certain degree of error. For an individual $n$ and an alternative $i$, it is assumed that there are indicators $I_{mni}$, $m=1,...,M$, which allow the identification of these constructs. There is a wide range of possible indicators that go from simple discrete responses regarding the perception of a variable to complex continuous variables.

ICLV models have been used to capture latent attributes such as product quality or individual characteristics such as attitudes, intentions, and beliefs (\cite{vij2016and}) and, recently, to capture instant utility in transport experiments, using physiological data as indicators (\cite{henriquez2025identifying}).

RUM can be easily interpreted, as modelers typically use linear additive functions, allowing to analytically derive values of economic and behavioral relevance, such as the willingness to pay for certain attributes. However, RUM models underperforms against data-driven AI or \textit{machine learning} (\textit{ML}) algorithms in terms of forecasting capabilities. AI algorithms, though, typically lack interpretability (\cite{arkoudi2023combining}).

Recently, several studies have proposed formulations that combine AI algorithms with typical discrete choice models, improving forecasting capabilities while maintaining a certain degree of interpretability (\cite{han2022neural,rodrigues2024model,aboutaleb2021discrete,arkoudi2023combining,sifringer2020enhancing}). For example \textcite{sifringer2020enhancing} proposed a formulation in which the systematic utility, for a given individual $n$ and an a given alternative $i$, can be written as follows:
\begin{equation}
V_{ni}=f_i(X_{ni},\beta)+r_i(D_n,\omega)
\end{equation}
where $f_i(X_{ni},\beta)$, the knowledge-driven part regarding alternative $i$, can be interpreted based on the relation between the attributes $X_ni$ and suited parameters $\beta$, while $r_i(D_n,\omega)$, the data-driven part regarding alternative $i$, is a representation component learned using an ML algorithm (e.g. a \textit{neural network}) based on the variables $D_n$ and the parameters $\omega$ and does not require to be interpretable. 

Hybrid ML-RUM models have been used to process complex data on factors that have a role in decision making, but do not necessarily need to be interpreted. For example, transport choice models have used neural networks to incorporate the representation of alternatives' images in discrete choice experiments (\cite{rossetti2019explaining,ramirez2021measuring,van2025computer,van2024utility}) and to better capture interindividual taste parameter heterogeneity (\cite{han2022neural}). Videos, on the other hand, have not been used in this literature. \textcite{millar2021space} implemented cameras on participants' bicycles to record the surroundings; however, these data were not used since automated processing presented major challenges. Instead, they collected historic Mapillary street view data to extract urban features. However, videos can also be analyzed with trained neural networks (e.g. \cite{pashchenko2019deep}) to identify objects. A shortcoming of this approach is that neural networks only identify those objects already included in the training data set and cannot make semantic interpretations of visual information.

An alternative approach is to use pre-trained LLM algorithms, as they are flexible enough to identify urban features and interpret traffic conditions. For example, \textcite{llm_video_traffic} used \textit{generative pre-trained transformer chat}
(\textit{ChatGPT}) to process video data and extract detailed semantic information about traffic conditions, road characteristics and driver behaviors. Similarly, ChatGPT has also been used for natural language processing and sentiment analysis (\cite{julianto2023enhancing,kheiri2023sentimentgpt,pullanikkat2024utilizing}). \textcite{henriquez2025evidence} used ChatGPT to embed a sentiment analysis in a travel satisfaction model. ChatGPT, however, has not yet been used to embed video information in a behavioral model. 

Using LLMs for data interpretation, a model can be formulated following a structure similar to \textcite{sifringer2020enhancing}. From this perspective, $r_i(D_n,\omega)$ becomes a function of features extracted from visual information using a LLM as an interpreter. This mixture of models has not been previously explored, despite having multiple potentials: enhancing discrete choice models' behavioral explainability without the need of training neural networks, gaining semantical interpretation of complex contextual information that can be easily collected in practice, and estimating the model just using the maximum likelihood method. In this paper, we explore this potential next generation of hybrid RUM-ML models for the particular case of explanation of cycling behavior. This allows us to discuss the consistency of ChatGPT's semantic interpretations of urban environment extracted from videos and the observed traffic behavior. 



\section{Methods}\label{sec:methods}

This section describes the participants' recruitment process, the instruments used in the experiment, the experiment design, and the modeling approach. We remark that the study was approved by the Ethics Committee of the Faculty of Physical and Mathematical Sciences at Universidad de Chile. 

\subsection{Participants}

The study began with an open call through the faculty's internal networks and its official social media, offering an incentive of 15,000 CLPs ($\approx$ 15 USD) where prior cycling experience was not required, only the ability to ride a bicycle, recruiting a total of 46 participants. All signed an informed consent form detailing the scope of the measurement phase and their responsibilities as participants. The mean age was 27.8 years with a standard deviation of 7.6 years. 33\% participants identified as female and 67\% as male. The sample consisted of 36 students, 8 workers and 2 individuals who declared ``Other'' as their main occupation.

Each participant completed a questionnaire to characterize their cycling experience, as well as two standardized surveys to evaluate their psychological state: the \textit{Depression Anxiety Stress Scale} (\textit{DASS-21}) and the \textit{Positive and Negative Affect Schedule} (\textit{PANAS}). Among the participants, 28 were frequent cyclists (cycling weekly) and 18 were occasional cyclists (less than weekly). Additionally, 25 participants reported familiarity with the route area. We highlight that both PANAS and DASS-21 have been validated for the Chilean population, indicating that they are effective tools to measure emotional states in studies applied to this population (\cite{panas_chile, panas_chile_2, chilean_dass21}). 

\subsection{Apparatus}

To conduct this experiment, we used the following instruments: \textit{Empatica E4}, \textit{ContextINO}, and \textit{Raspberry Pi} devices, a geopositioning app, and a \textit{GoPro Hero 8} camera. The Empatica E4  is a wristband that measures physiological signals such as \textit{photoplethysmography} (\textit{PPG}), electrodermal activity (EDA), \textit{skin temperature} (\textit{SKT}), and motion data from an accelerometer and gyroscope (\cite{mccarthy2016validation}). The ContextINO is a set of environmental sensors including $CO_2$, noise, brightness, and humidity (\cite{barriaRelatingEmotionsPsychophysiological2023}). 

The GoPro camera was mounted on the bicycle's handlebar to capture visual context during the ride. The ContextINO device, along with the Raspberry Pi and its battery, was placed on the bicycle's rear rack inside a specially designed polyurethane casing. This setup ensured that the sensors for CO\textsubscript{2}, ambient light, and noise were adequately exposed to the environment. The Empatica E4 was synchronized with a mobile phone that also displayed the route using the GPS app \textit{GuruMaps}. The mobile phone, secured to the handlebar with a mount, facilitated the participant's navigation and recorded the GPS data, essential for correlating physiological and contextual data. Safety equipment, such as helmets and gloves, was provided to each participant to ensure their well-being during the experiment. The combination of these instruments allowed for the simultaneous collection of physiological signals, environmental context, and positional data, critical for the analysis factors that affect cyclists' arousal and fatigue states in real-world urban settings.

\begin{figure}
    \centering
    \includegraphics[width=10cm]{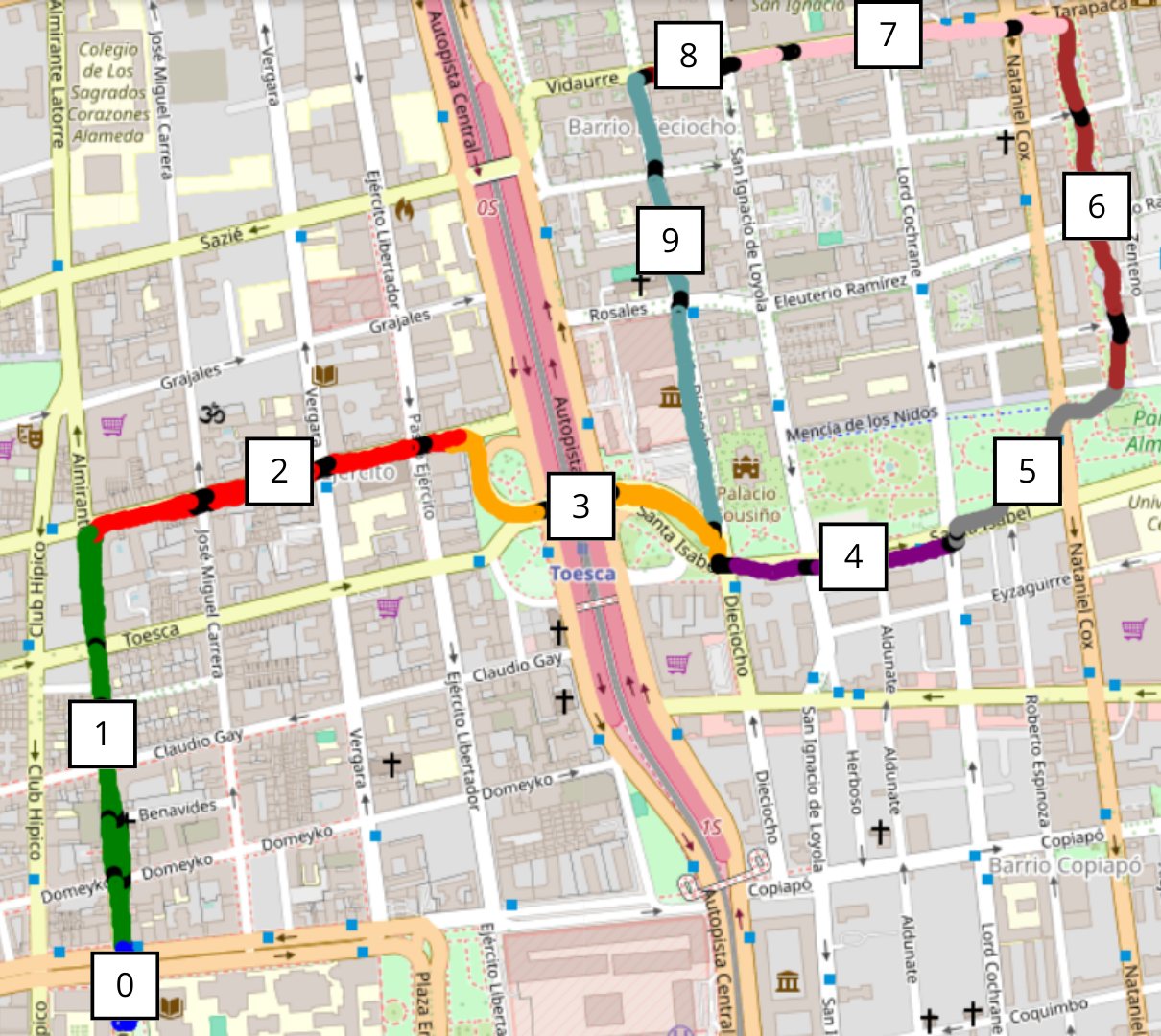}
    \caption{Cycling study route.}
    \label{route}
\end{figure}

Figure~\ref{route} shows the route taken by the participants. It started and ended at the Faculty of Physical and Mathematical Sciences of Universidad de Chile and passed through various points in downtown Santiago, including relevant urban locations such as the cross of the central highway, \textit{Parque Almagro}, and \textit{Paseo Bulnes}. The route was divided into nine distinct segments, each with invariant infrastructure characteristics. For each segment, the following attributes were recorded: type of infrastructure (including streets with bike lanes, streets without bike lanes, and parks or promenades); the width of the bike lane (in meters); the vehicular capacity of the street (number of automobile lanes); and direction (whether the cyclist was moving with or against vehicular traffic). These data were included in the model as infrastructure features.

\subsection{Experimental Design}

The experimental design considered a protocol prior to the cycling experience. The study involved 46 participants who were asked to complete a physical health survey beforehand. Upon arrival, their identification data was recorded, important time milestones were noted, blood pressure was measured, and an introduction to the study was provided. Participants then completed the PANAS and DASS21 standardized surveys, and a resting baseline of psychophysiological signals was captured using the \textit{Empatica} wristband; this period was referred to as ``baseline 1''. Before starting the cycling experience, it was ensured that all devices were functioning correctly and had sufficient battery life. Participants tested the bicycle before the trip to familiarize themselves with the brakes and adjust the seat; this period was referred to as ``baseline 2''. Then the \textit{ContextINO} device was turned on, the recording in \textit{GuruMaps} was started, and the GoPro camera was activated. Once the camera was recording, the screen of a mobile phone displaying the official time in \texttt{hh:mm:ss} format was shown to synchronize the start of the experience. After these preparations, participants began their trip and the start time was recorded. During the ride, an audio file played through headphones prompted the user every three minutes with the following question: "Which of the following emotions best describes you right now? Happy, relaxed, bored, or stressed?" These emotion labels are based on the classification used in the \textit{circumplex model of affect}, adapted by (\cite{barriaRelatingEmotionsPsychophysiological2023}). The verbal responses were then recorded with the GoPro camera. Participants were also encouraged to report their emotions spontaneously at any point during the ride, not just when prompted. At the end of the ride, they turned off their cameras, stopped location tracking, and met with the study coordinator to return their equipment.

\subsection{Data Processing}

\textbf{Behavioral variables}

To determine the dependent variable, the cyclist's action was imputed for every 5-second time window. An algorithm developed specifically for this study classified actions into five categories: maintaining speed, accelerating, decelerating, braking, or waiting. This algorithm uses changes in speed, derived from GPS data, and a threshold-based system to assign the actions. Details can be found in the Appendix \ref{appendix:algoritmo_acciones}.

\noindent\textbf{Psychophysiological Data Processing}

To process PPG signals, the \textit{NeuroKit2} library (\cite{Makowski2021}) in \textit{Python} was used. The procedure began with the identification of signal interruptions, removing all segments of the dataset where no data were recorded for more than 5 seconds. This prevents the inclusion of defective segments that could negatively affect the detection of RR intervals (i.e., the time between successive heartbeats). The signal was then resampled to a fixed frequency of 45 Hz. The \texttt{nk.ppg\_process()} function from \textit{NeuroKit2} was then used to process the PPG signals and calculate the \textit{heart rate} (\textit{HR}), the \textit{standard deviation of normal-to-normal} intervals (\textit{SDNN}), and the \textit{root mean square of successive differences} (\textit{RMSSD}). Finally, the extracted features were aggregated into time windows to facilitate analysis.

Due to the high level of noise present in PPG signals collected outside laboratory conditions, \textit{Kubios} was used as an alternative processing method. This is a specialized software for processing cardiovascular activity signals such as PPG and \textit{electrocardiogram} (\textit{ECG}) (\cite{TARVAINEN2014210}). \textit{Kubios} provides advanced tools for analyzing and filtering physiological signals and has been widely used to process PPG data, especially when collected in uncontrolled environments with motion-related \textit{artifacts} (\cite{Milstein8, Schuurmans2020}).

To process the EDA signals, \textit{NeuroKit2} (\cite{Makowski2021}) was used, a library that offers tools specifically designed for this type of signal. All segments with no recorded data for more than 5 seconds were removed, so defective segments were not included. Subsequently, the EDA signal was resampled to a frequency of 8 Hz. Prior to applying \textit{NeuroKit2} processing, three data-cleaning steps were performed: first, a first-order \textit{Butterworth bandpass} filter was applied to eliminate frequencies outside the range of interest, smoothing the signal and removing high-frequency noise; then, outliers were handled through an \textit{interpolation-based outlier} filter, ensuring signal continuity and reducing distortion caused by artifacts (i.e., non-physiological signal distortions such as motion); finally, the \textit{Haar Wavelet transform} was used to remove remaining motion artifacts. After this preprocessing, the \texttt{nk.eda\_process()} function from \textit{NeuroKit2} was used to decompose the EDA signal into its tonic and phasic components and identify \textit{skin conductance response}s (\textit{SCR}s) along with their parameters: \textit{Amplitude}, \textit{RiseTime}, and \textit{RecoveryTime}. Finally, the extracted features were aggregated into specific time windows, calculating various statistical parameters such as standard deviation, mode, and median for use in the models. We remark that the \texttt{nk.eda\_process()} function performs a detection process for SCRs, which is highly sensitive to motion artifacts. Thus, several filters are applied before using this function, in order to remove noise and ensure that the resulting parameters are of the highest possible quality. 

\noindent\textbf{Large Languague Model-based Video Descriptor}

\begin{figure}[H]
    \centering
    \includegraphics[width=15cm]{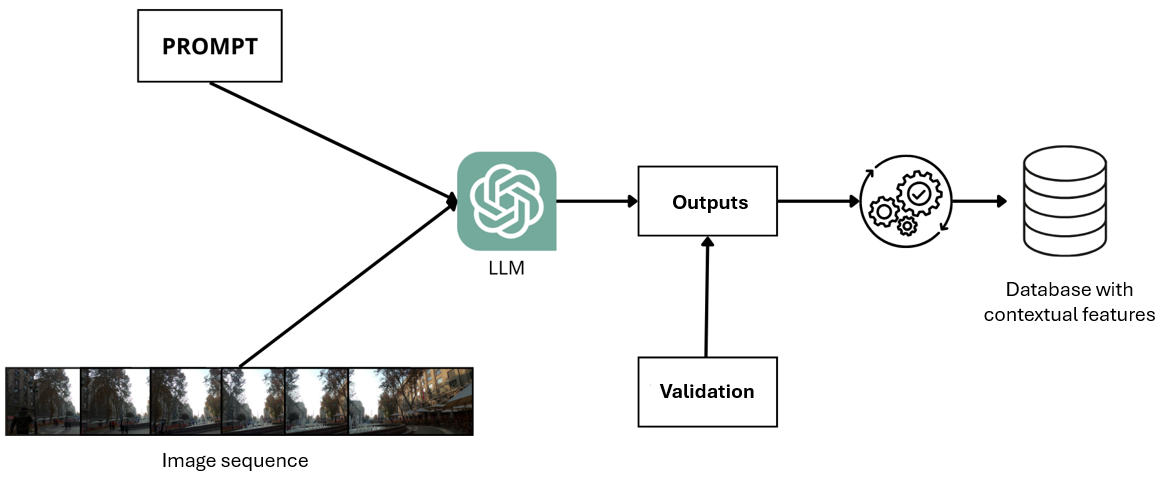}
    \caption{LLM Video Descriptor (LVD) process.}
    \label{fig:LVD}
\end{figure}

To implement the strategy discussed in the theoretical framework, we developed a process called the \textit{LLM Video Descriptor} (\textit{LVD}), which utilizes LLMs with multimodal capabilities to generate textual descriptions of urban environments from video data. Based on the tools available as of August 2024, we opted for the \textit{GPT-4} model through the \textit{OpenAI API} (accessible at \href{https://openai.com/}{https://openai.com/}). The LVD process (\autoref{fig:LVD}) begins by creating a prompt and selecting image sequences extracted from video recordings, both included in a \textit{JSON} file along with other necessary parameters (such as user credentials). An output text is then processed to extract a database in which each column collects a variable of interest. This process includes two key validation stages: an initial validation in which different prompts are tested and adjusted before scaling the process, and a subsequent validation in which the consistency of the model for generating environmental variables is evaluated. As these services mean a cost based on tokens used (both input, such as images and prompts, and output, the generated description), it is crucial to optimize the prompt and the selection of images to achieve an efficient analysis in terms of cost and performance.

After the initial validation and multiple iterations, we obtain the final prompt (Appendix~\ref{appendix:gpt_prompt}). Then, we process all available image sequences, obtaining a database with 12 georeferenced variables describing the urban environment. These include, among others, proximity to different types of vehicle (e.g., buses, trucks), proximity of pedestrians (children, adults, groups of people, pets), identification of red traffic lights and indicators of infrastructure quality (Appendix \ref{appendix:gpt_results}). 

Note that the 12 urban environment descriptors obtained were not identified by directly prompting the LLM to map them from video data but were selected by the LLM own potential variables of interest. The process began by prompting the LLM to provide a semantic description, specifically asking it to assign a categorical stress level that a cyclist could experience in each context, based on its expert judgment as an evaluator of cycling conditions and infrastructure (prompt in Appendix~\ref{appendix:gpt_prompt}). The LLM then generates a textual explanation for the assigned stress levels, from which the relevant features are extracted. These semantic and subjective variables are especially valuable, as they not only characterize the environment but also enable us to assess and validate the LLM’s decision-making criteria. Although we apply this approach to cycling route analysis, its flexibility makes it suitable for other applications requiring customized, image-derived descriptions. 

\begin{figure}[H]
	\centering
	\begin{subfigure}[b]{0.44\textwidth}
		\centering
		\includegraphics[width=\textwidth]{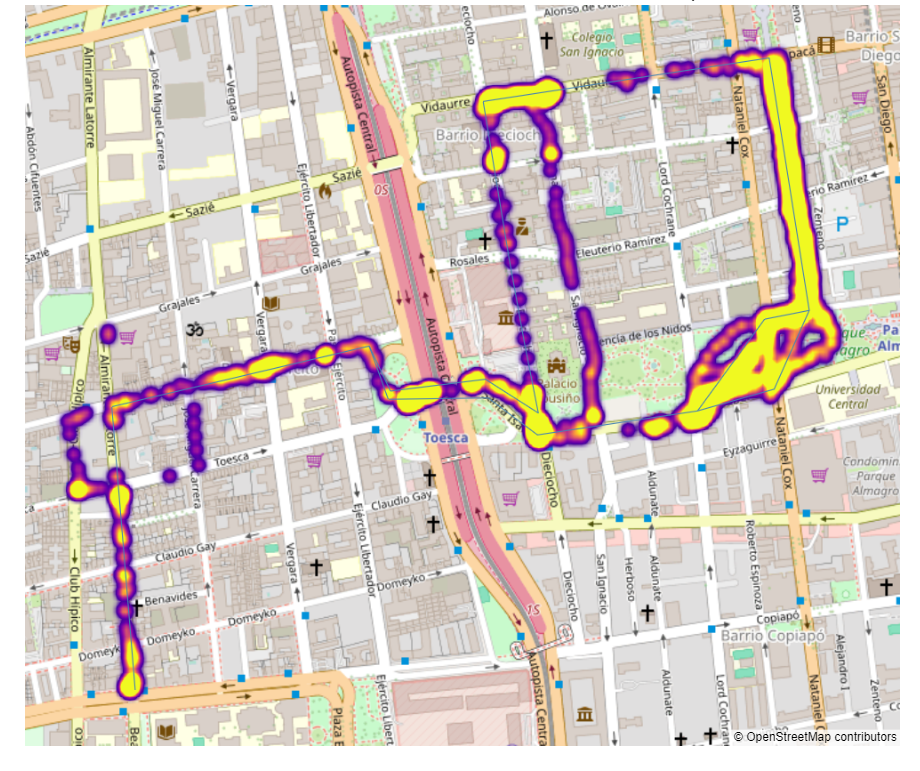}
		\caption{Poor Cycling Infrastructure.}
		\label{fig:badinfra}
	\end{subfigure}
	\hfill
	\begin{subfigure}[b]{0.44\textwidth}
		\centering
		\includegraphics[width=\textwidth]{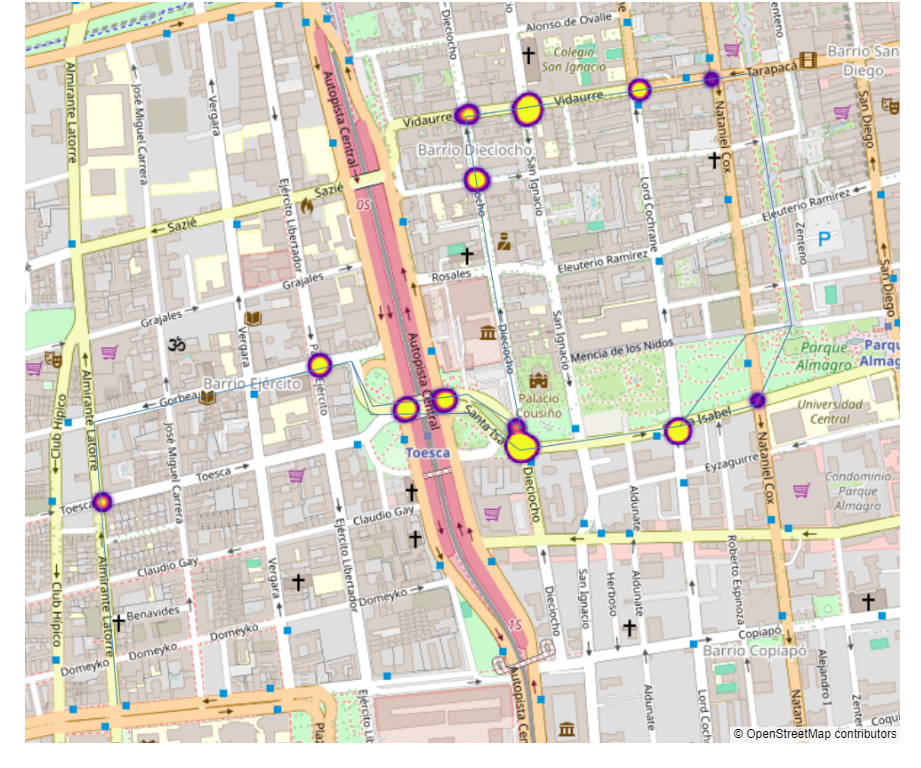}
		\caption{Red Traffic Lights}
		\label{fig:Redlights}
	\end{subfigure}
    \caption{Heatmaps of semantic causes of cyclist's stress generated by LLM Video Descriptor (LVD).}
    \label{fig:LVMexample}
\end{figure}

\autoref{fig:LVMexample} shows an example of two LVD variables, identified by the LLM as the cause of the stress perceived by the cyclist, aggregated over all individuals in the sample. LDV explanations for stress are, on the left, bad infrastructure and, on the right, red traffic lights. Note that figure includes data from a subject who took a wrong turn during the experiment. Appendix \ref{appendix:gpt_results} shows the details of all features identified through the LVD.

\begin{figure}[ht!]
	\centering
	\includegraphics[width=1\linewidth]{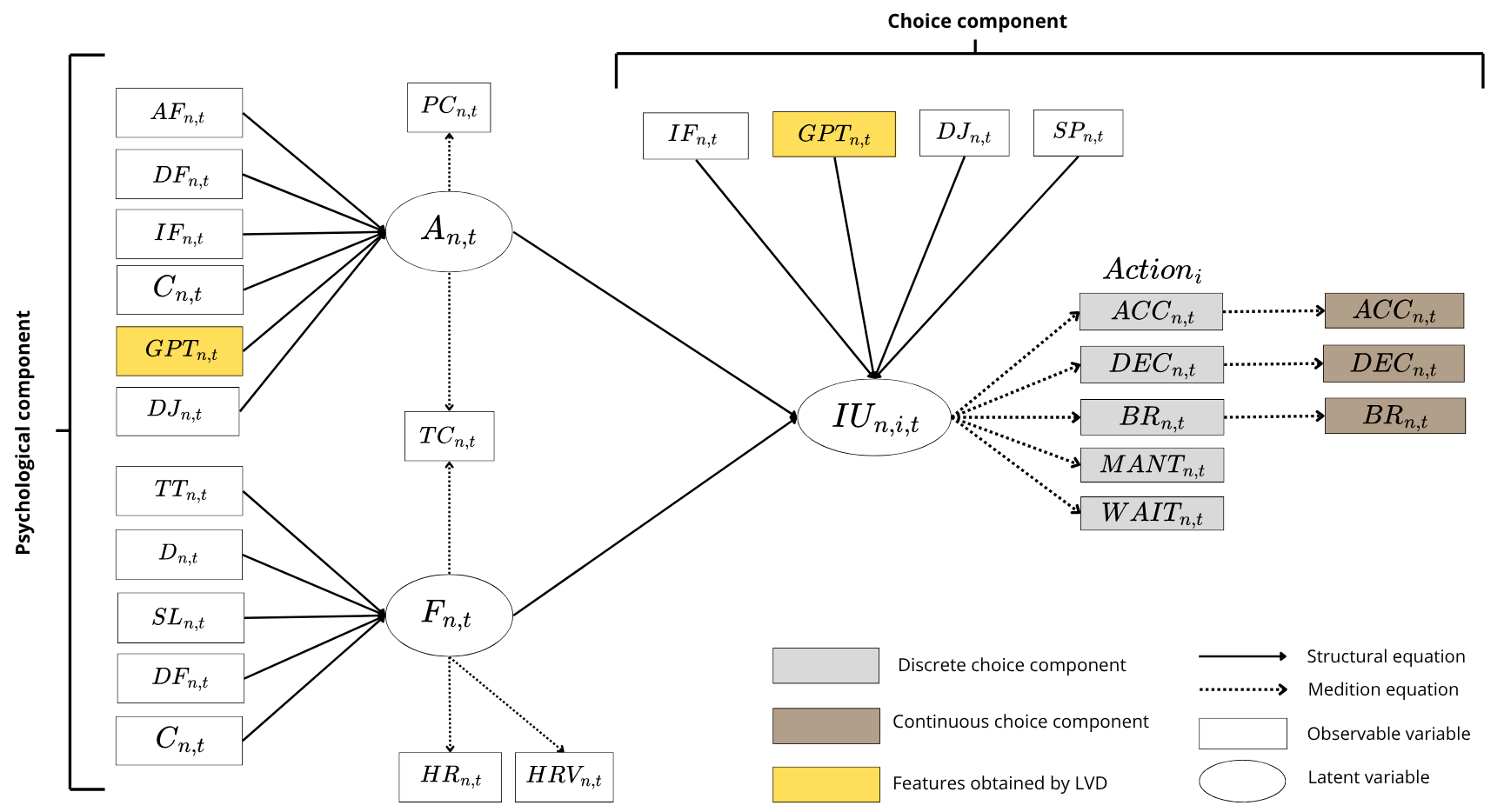}
	\caption{Diagram of the proposed ICLV model. Observable variables in rectangles; latent variables in ellipses.}
	\label{iclv_model}
\end{figure}

\noindent\textbf{Integrated Choice and Latent Variable Model for Cyclist Action}\\
This section presents the details of the proposed ICLV model (\autoref{iclv_model}) based on the VR study conducted by \cite{bogaczModellingRiskPerception2021a}, which modeled these choices as a function of latent risk perception measured by \textit{electroencephalography} (\textit{EEG}).

In our model, we assume that individuals choose among a set of five cycling actions at each time instant: accelerating, braking, decelerating, maintaining speed, and waiting. Each alternative has a level of \textit{instant utility} i.e. the perceived level of satisfaction at each instant during an experience (\cite{kaneman}), following \textcite{henriquez2025identifying}. Thus, the action with the maximum instant utility is chosen. In addition, instant utilities are influenced by two latent variables: \textit{arousal} and \textit{fatigue}, representing states of mental and physical stress, respectively.
    
    
First, we describe the structural equations, followed by the instant utility functions for each action. Then, the measurement equations are introduced, along with the density function that defines the probability of selecting an action, and the density corresponding to the continuous component of the choices. Finally, the likelihood function of the complete model is addressed.

\noindent\textbf{Structural Equations}\\
The model includes two latent components to capture the impact of fatigue and arousal on cyclist actions. 

The \textit{latent fatigue} (Eq. \ref{fatigaeq}), for individual $n$ at instant $t$, depends on: \textit{elapsed travel time} $TT_{n,t}$, \textit{traveled distance} $D_{n,t}$, \textit{slope} $SL_{n,t}$, \textit{demographic features} \textit{DF}$_{n,t}$, \textit{ContextINO variables} \textit{C}$_{n,t}$, and a noise $\varepsilon^{fat}_{n,t} \sim N(0,1)$. $\beta_0^{fat}$, $\beta_{TT}$, $\beta_{DR}$, $\beta_{SL}$, $\beta_{DF}$, and $\beta_{C}$ are suited parameters for the latent fatigue function.
\begin{equation}
	\label{fatigaeq}
	F_{n,t} = \beta_0^{fat} + \beta_{TT} \cdot TT_{n,t} + \beta_{DR} \cdot D_{n,t} + \beta_{SL} \cdot SL_{n,t} + \beta_{DF} \cdot DF_{n,t} + \beta_{C} \cdot C_{n,t} + \varepsilon_{n,t}^{fat}
\end{equation}

The \textit{latent arousal} (Eq. \ref{activaeq}), for individual $n$ at instant $t$, depends on: \textit{audio features} $AF_{n,t}$, \textit{demographic features} $DF_{n,t}$, \textit{infrastructure features} $IF_{n,t}$, \textit{ContextINO variables} $C_{n,t}$, \textit{GPT environment variables} $GPT_{n,t}$, \textit{distance to junction} \textit{DJ}$_{n,t}$, and a noise $\varepsilon^{act}_{n,t} \sim N(0,1)$. $\beta_0^{act}$, $\beta_{AF}$, $\beta_{DF}$, $\beta_{IF}$, $\beta_{C}$, $\beta_{GPT}$, and $\beta_{DI}$ are suited parameters for the latent arousal function.
\begin{equation}
	\label{activaeq}
	A_{n,t} = \beta_0^{act} + \beta_{AF} \cdot AF_{n,t} + \beta_{DF} \cdot DF_{n,t} + \beta_{IF} \cdot IF_{n,t} + \beta_{C} \cdot C_{n,t} + \beta_{GPT} \cdot GPT_{n,t} + \beta_{DJ} \cdot DJ_{n,t} + \varepsilon_{n,t}^{act}.
\end{equation}

To capture nonlinear aspects of cyclist behavior related to speed and distance to the intersection, two levels were defined: \textit{low} and \textit{high} for both variables. These levels were established using the quartiles of their respective distributions, where \textit{low} corresponds to values below the first quartile and \textit{high} to values above the third quartile. For distance, the low level was defined as values below \(9.7\) [m], and high as values above \(82.8\) [m]. For speed, the thresholds were \(4.4\) [km/h] for low and \(17.4\) [km/h] for high.

Infrastructure features were determined using a segment assignment method, allowing the identification of cycling lanes, road width, vehicle direction relative (to cyclist), and number of lanes. Contextual variables were derived from the LVD process, to build a suitable environment able to explain cyclists' decisions. 

Equations \eqref{ut1} to \eqref{ut5} present the instant utility functions of the actions that cyclist $n$ could choose at a time instant $t$: accelerating ($i=1$), braking ($i=2$), decelerating ($i=3$), waiting ($i=4$), and maintaining speed ($i=5$). All of them depend on the cyclist's distance to the intersection, speed, infrastructure features, and contextual variables derived from LVD. Maintaining speed is used as a reference for the rest of the actions, as it is a passive action indicating an equilibrium state for the cyclist.
\begin{equation}
	\label{ut1}
	\begin{gathered}
		IU_{n,t}^{1} = \beta_0^{1} + \beta_{DJ}^{1} \cdot DJ_{n,t} + \beta_{IF}^{1} \cdot IF_{n,t} + \beta_{A}^{1} \cdot A_{n,t} + \beta_{F}^{1} \cdot F_{n,t} + \beta_{GPT}^{1} \cdot GPT_{n,t} +
		\beta_{SP}^{1} \cdot SP_{n,t} + \varepsilon_{n,t}^{1}
	\end{gathered}
\end{equation}
\begin{equation}
	\label{ut2}
	\begin{gathered}
		IU_{n,t}^{2} = \beta_0^{2} + \beta_{DJ}^{2} \cdot DJ_{n,t} + \beta_{IF}^{2} \cdot IF_{n,t} + \beta_{A}^{2} \cdot A_{n,t} + \beta_{F}^{2} \cdot F_{n,t} + \beta_{GPT}^{2} \cdot GPT_{n,t} +
		\beta_{SP}^{2} \cdot SP_{n,t} + \varepsilon_{n,t}^{2}
	\end{gathered}
\end{equation}
\begin{equation}
	\label{ut3}
	\begin{gathered}
		IU_{n,t}^{3} = \beta_0^{3} + \beta_{DJ}^{3} \cdot DJ_{n,t} + \beta_{IF}^{3} \cdot IF_{n,t} + \beta_{A}^{3} \cdot A_{n,t} + \beta_{F}^{3} \cdot F_{n,t} + \beta_{GPT}^{3} \cdot GPT_{n,t} +
		\beta_{SP}^{3} \cdot SP_{n,t} + \varepsilon_{n,t}^{3}
	\end{gathered}
\end{equation}
\begin{equation}
	\label{ut4}
	\begin{aligned}
		IU_{n,t}^{4} &= \beta_0^{4} + \beta_{DJ}^{4} \cdot DJ_{n,t} + \beta_{IF}^{4} \cdot IF_{n,t} + \beta_{A}^{4} \cdot A_{n,t} + \beta_{F}^{4} \cdot F_{n,t} + \beta_{GPT}^{4} \cdot GPT_{n,t} + \varepsilon_{n,t}^{4}
	\end{aligned}
\end{equation}
\begin{equation}
	\label{ut5}
	IU_{n,t}^{5} = \varepsilon_{n,t}^{5}
\end{equation}
In Equations \eqref{ut1} to \eqref{ut5}, in addition to the already defined terms, $SP_{n,t}$ is the cyclist’s speed. All $\beta_{*}^{*}$ are suited constants, and all  $\varepsilon^{*}_{*}$ are Gumbel errors $\sim N(0,1)$.

All cases in which participants stopped at a signalized intersection were excluded, as they were imposed to stop. This particular case is reflected in Equation \ref{ut5}, which depends only on the random error term. In the action imputation algorithm, the assignment of this action depends exclusively on speed, which could introduce a spurious relationship. For this reason, Equation \eqref{ut4} does not include speed-related parameters.

\noindent\textbf{Measurement Equations}\\
The final component of the ICLV models is the measurement equations, used to link the latent variables with their observable indicators. They relate arousal and fatigue to specific physiological and contextual measures.

Arousal is measured by the \textit{tonic component} (\textit{TC}) and \textit{phasic component} (\textit{PC}) of the EDA signal. TC reflects the background electrodermal activity associated with the general arousal state, representing the accumulated stress or effort. PC captures rapid and transient responses to specific stimuli, associated with events that could be annoying or disruptive for cyclists. This dual approach captures both the short- and long-term effects of the urban environment on cyclists' states.

To measure fatigue, we use the \textit{heart rate} (\textit{HR}) and the \textit{heart rate variability} (\textit{HRV}). HR is a direct indicator of physical load, as its increase reflects greater physical effort, common in high-demand situations in urban cycling, while HRV depicts the evolution of this physical effort.

Equations \ref{toniceq}, \ref{phasiceq}, \ref{hreq}, and \ref{hrveq} describe how arousal and fatigue manifest through observed indicators. In the model, these indicators are assumed to be normally distributed around an expected value that depends only on the latent variables, except for TC which, as shown in equation \ref{toniceq}, depends on both latent variables. This allows control for the effect of physical effort on TC, as sweat levels significantly affect the signal. For each indicator, there are factors $\gamma_{*}$ for both latent variables, and there are errors $\varepsilon_{*}$, assumed to be normally distributed with mean 0 and standard deviation $\sigma_{\eta}$.
\begin{equation}
	\label{toniceq}
	TC_{n,t} = \gamma_{TC} \cdot A_{n,t} + \gamma_{TC} \cdot F_{n,t} + \varepsilon_{TC}
\end{equation}
\begin{equation}
\label{phasiceq}
	PC_{n,t} = \gamma_{PC} \cdot A_{n,t} + \varepsilon_{PC}
\end{equation}
\begin{equation}
\label{hreq}
	HR_{n,t} = \gamma_{HR} \cdot F_{n,t} + \varepsilon_{HR},
\end{equation}
\begin{equation}
\label{hrveq}
	HRV_{n,t} = \gamma_{HRV} \cdot F_{n,t} + \varepsilon_{HRV}
\end{equation}

\noindent\textbf{Action Probability}\\
The action alternatives, accelerating, braking, decelerating, maintaining speed, and waiting, are all linked to speed changes: accelerating refers to its increase, braking involves its abrupt decrease, and decelerating means reducing it gradually. In our formulation, the speed of an individual $n$ at a time $t$ is denoted as $x_{n,t}$.

This distinction was made to account for the different underlying reasons for these actions. For example: braking may result from the sudden appearance of a car at an intersection, which increases risk collision; decelerating may be chosen when there is no imminent risk but a need for caution (e.g., the cyclist is still far from the intersection) or when the cyclist stops pedaling due to physical fatigue. Cases where a person moves at a constant speed are classified as maintaining speed. Finally, waiting refers to cases where a cyclist has stopped and remains stationary. 

The probability that individual $b$ chooses to perform action $i$ at time $t$ is given by Equation \ref{prob_ac}:
\begin{equation}
	\label{prob_ac}
	P\left( y_{n,t,i} = 1 \mid \mathbf{x}_{n,t}, A_{n,t}, F_{n,t} \right) = \frac{\exp\left(IU_{n,t}^{i}\right)}{\sum_{j=1}^{5} \exp\left(IU_{n,t}^{j}\right)}
\end{equation}
Note that this probability expression corresponds to a logit function with a dispersion parameter equal to 1.

\noindent\textbf{Continuous Choice Component}\\
The continuous component provides additional details about cyclist's behavior by indicating the magnitude of the chosen action. It is not meant to improve model efficiency, but rather to enhance behavioral understanding. The dependent variables are the continuous values of acceleration, braking, and deceleration, only evaluated if those actions were selected in the discrete component.

To model the magnitude, a normal distribution is assumed, conditional on the cyclist's speed. Equation \eqref{cont_comp} expresses the likelihood—that is, the probability density—of observing the continuous component of individual $n$'s action $i$ at time $t$, given the covariates $x_{n,t}$, with expected value $\mu_i$ and standard deviation $\sigma_i$:
\begin{equation}
	\label{cont_comp}
	P\left( y_{n,t,i}^{cont} \mid x_{n,t} \right) = \frac{1}{\sigma_{i} \sqrt{2\pi}} \exp\left( -\frac{\left(y_{n,t,i}^{cont} -  x_{n,t}  \mu_{i} \right)^2}{2\sigma_{i}^2} \right)
\end{equation}

\noindent \textbf{Likelihood}\\
The likelihood is modeled considering both discrete and continuous choices, integrating over the measurement error terms $\eta$. Equation \ref{likelihood} shows the total likelihood $LL$ (conditional terms omitted for simplicity).
\begin{equation}
	\label{likelihood}
	LL = \prod_{n=1}^{N} \left( \prod_{t=1}^{T_n} \int_{\eta_{med}} \left( \prod_{i=1}^{J} P\left( y_{nti} \mid A_{nt}, F_{nt}, \mathbf{x}_{nt} \right)^{y_{nti}} \times P\left( y_{n,t,i}^{cont} \mid x_{n,t} \right)^{\delta_{\text{cont}}(i)} \right) f(\eta_{med}) d\eta_{med} \right)
\end{equation}

The term $P\left( y_{nti} \mid A_{nt}, F_{nt}, \mathbf{x}_{nt} \right)^{y_{nti}}$ denotes the probability that cyclist $n$ chooses action $i$ at time $t$, raised to the power $y_{nti}$ to indicate selection. $\delta_{\text{cont}}(i)$ is an indicator equal to 1 if the action has a continuous component. The integral is performed over measurement errors that are assumed to be normally distributed.

\noindent\textbf{Simulated Likelihood}\\
The simulated likelihood $LL_{sim}$ approximates the integral in equation \ref{likelihood} with an average over $R$ simulations. Each simulation, from $r=1,...,R$, draws a value $\eta_{med}^{(r)}$ from the error distribution. Equation \ref{likelihood_simulated} expresses the obtained approximation.
\begin{equation}
	\label{likelihood_simulated}
	LL \approx \prod_{n=1}^{N} \left( \prod_{t=1}^{T_n} \frac{1}{R} \sum_{r=1}^{R} \left( \prod_{i=1}^{J} P\left( y_{n,t,i} \mid \eta_{med}^{(r)} \right)^{y_{n,t,i}} \times P\left( y_{n,t,i}^{\text{cont}} \mid \eta_{med}^{(r)} \right)^{\delta_{\text{cont}}(i)} \right) \right)
\end{equation}

\noindent\textbf{Models for Comparison}

To assess the effects of including latent variables and multimodal data fusion, two versions of a \textit{multimodal logit} (\textit{MNL}) model and four versions of the proposed ICLV model are compared. Each version differs in the inclusion of variables from audio, image, or video data obtained through LVD. The two versions of MNL are a basal \textit{MNL} and one that incorporates explanatory variables from GPT \textit{MNL (I)}. The ICLV versions are all hybrid models: basal \textit{HM}, with audio features \textit{HM (A)}, with image features \textit{HM (I)}, and with audio and image features \textit{HM (IA)}.

\begin{figure}[H]
	\centering
	\includegraphics[width=10cm]{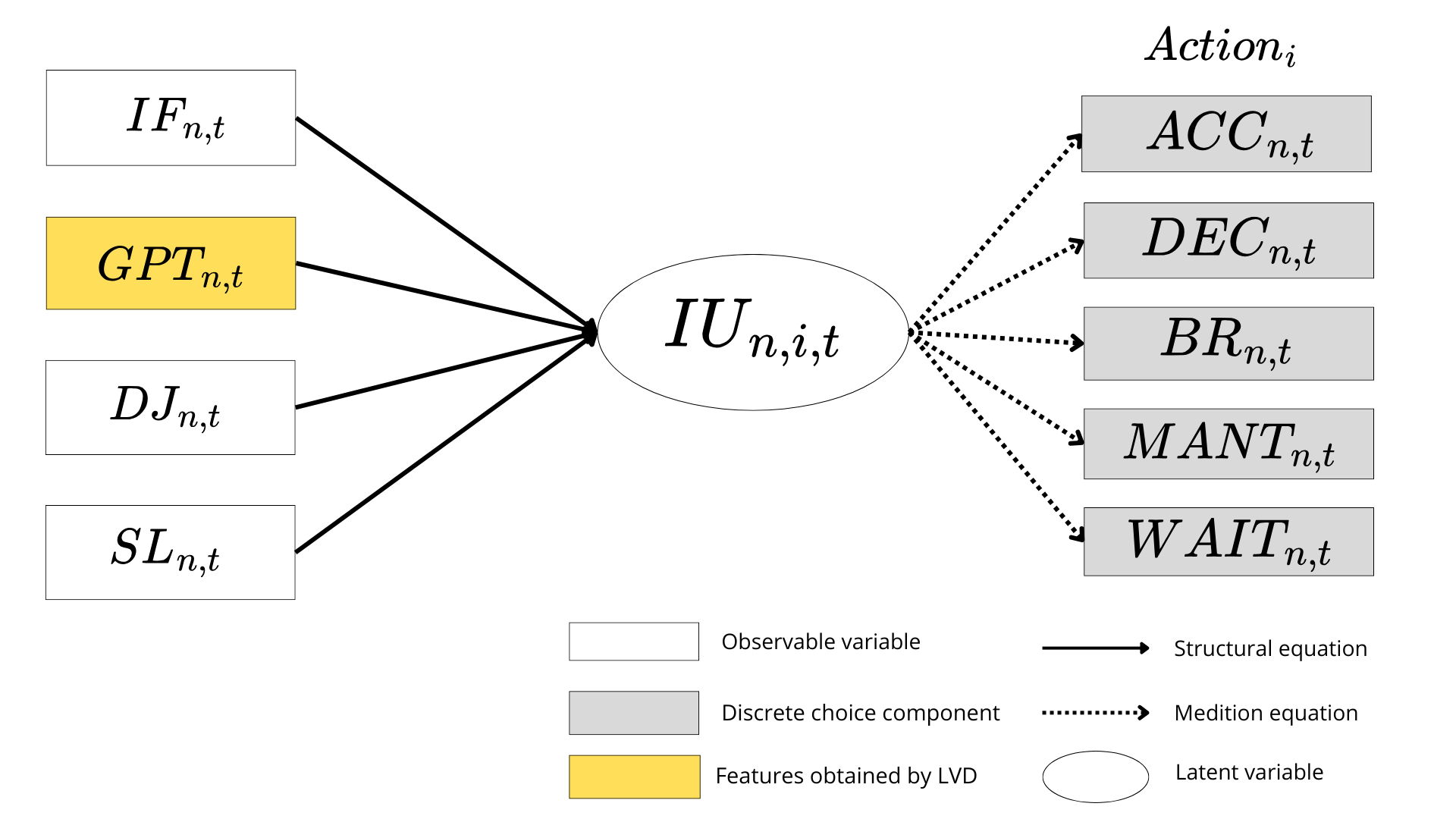}
	\caption{MNL model diagram.}
	\label{mnl_model}
\end{figure}

The base MNL model (Figure \ref{mnl_model}) does not include latent psychological variables or a continuous component, but does consider cyclists' baseline psychological states from surveys. It shares the discrete choice component with the ICLV but excludes latent variables.



\section{Results}\label{sec:results}

The Apollo library (\cite{Hess2022}), available in R (\cite{R2023}), was used to estimate the models. Before defining the final set of explanatory and measurement variables, several preliminary tests were conducted. These tests revealed that the mean value of EDA's tonic component TC and the standard deviation of its phasic component are the best variables to measure activation. Regarding fatigue, it was determined that RMSSD is the most effective variable for measuring HRV.

Models were estimated using the \texttt{apollo\_searchStart} function, designed to perform a global search for optimal parameters' starting points, minimizing the risk of converging on local maxima of the likelihood function. 
The estimation of all models was carried out using the infrastructure of the \textit{National Laboratory for High Performance Computing} (\textit{NLHPC}) at Universidad de Chile, and took nearly 8 days to complete. 

Table \ref{tab:model_metrics} presents the metrics summarizing the performance of each model. As shown, the hybrid models (HM) or ICLVs outperform the classic discrete choice models (MNL) in terms of fit metrics such as the \textit{final log-likelihood} (\textit{LL(final)}), the \textit{Akaike Information Criterion} (\textit{AIC}), the \textit{Bayesian Information Criterion} (\textit{BIC}), and the \textit{adjusted Rho-squared}. This result indicates that incorporating latent variables in ICLVs enhances the model’s ability to capture cyclist behavior. Moreover, the inclusion of image-derived data in the models, both in MNL and ICLV, yields additional improvements in performance metrics, suggesting that images provide valuable information that helps to better explain cyclist decisions in urban environments.
However, the effect of audio is not as significant as that of images. 
\begin{table}[ht!]
	\centering
	\caption{Model results, comparative metrics.}
	\label{tab:model_metrics}
	\resizebox{\textwidth}{!}{%
		\begin{tabular}{lccccccc}
			\hline
			\textbf{Model} & \textbf{N° parameters} & \textbf{LL(0)} & \textbf{LL(final)} & \textbf{AIC} & \textbf{BIC} & \textbf{Adj. \(\boldsymbol{Rho^2}\)} \\			 
			\hline
			MNL & 35 & -12584.2 & -10745.71 & 21561.4 & 21805.2 & 0.1433 \\
			MNL(I) & 47 & -12584.2 & -10516.50 & 21127.0 & 21454.3 & 0.1606 \\
			HM & 82 & -12584.2 & -10646.13 & 123730.3 & 124301.4 & 0.1475 \\
			HM(I) & 99 & -12584.2 & -10412.92 & 123191.8 & 123881.3 & 0.1647 \\
			HM(A) & 86 & -12584.2 & -10674.30 & 123573.8 & 124172.8 & 0.1449 \\
			HM(IA) & 103 & -12584.2 & -10442.41 & 123101.9 & 123819.2 & 0.1620 \\
			
			\hline
		\end{tabular}%
	}
\end{table}

Table \ref{tab:choice_component}, in Appendix \ref{appendixtableresults}, presents the results of the choice components for the six models compared. Each column represents a different model, while each row includes the estimated parameters for an explanatory variable and a specific action, along with their corresponding \textit{robust t-test}. 

\begin{figure}[H]
    \centering
    \includegraphics[width=14cm]{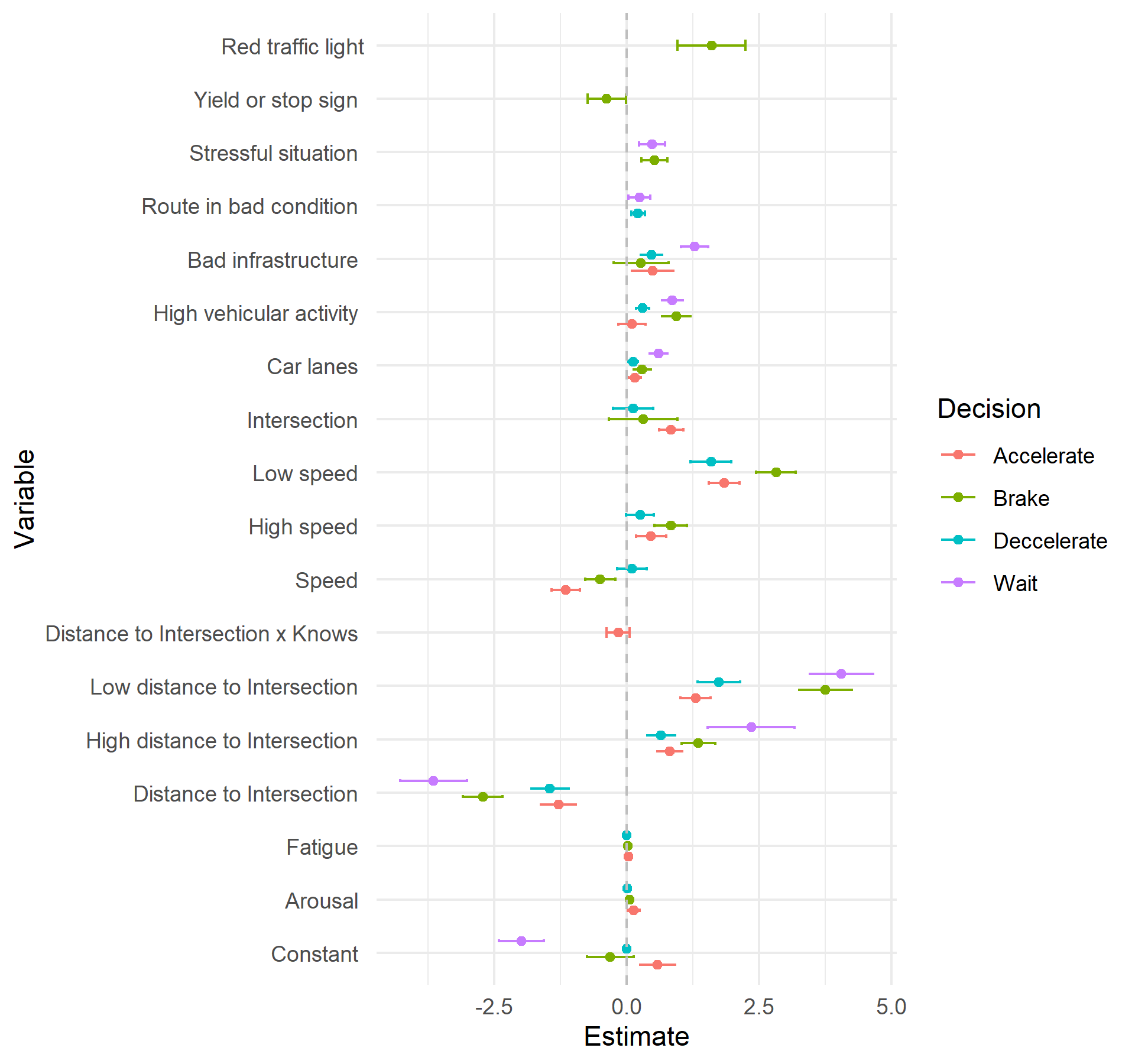}
    \caption{HM(IA) estimates and 90\% confidence interval.}
    \label{fig:output}
\end{figure}

\autoref{fig:output} summarizes the estimates of the most complex model: HM(IA). The results suggest that when the distance to the intersection is high (\textit{Dist. to intersection}), the probability of \textit{maintaining speed} is the highest, while the options of \textit{braking} and \textit{waiting} are significantly less likely. On the other hand, when the distance to the intersection is short, the probability of \textit{maintaining speed} drops sharply, while the probabilities of \textit{braking} and \textit{waiting} increase. This suggests that intersections have a significant effect, causing cyclists to slow down or stop. These conclusions are consistent in the models \textit{MNL(I)}, \textit{HM(A)}, and \textit{HM(IA)}.

Regarding the analysis of the \textit{speed} variable, it appears that at higher speeds the most probable actions are \textit{decelerating} and \textit{maintaining speed}. These are followed by \textit{braking} and \textit{accelerating}. This may indicate that cyclists prefer to maintain their speed within certain limits that allow them to feel comfortable or safe, as they significantly prefer to maintain or slightly reduce their speeds rather than increase them. 

When the speed is low (below 4 km/h), the probability of \textit{braking} increases drastically. This makes sense since, when the speed is low, cyclists are typically in situations that require them to brake to a stop. This behavior holds across all models. 

\begin{figure}
    \centering
    \includegraphics[width=15cm]{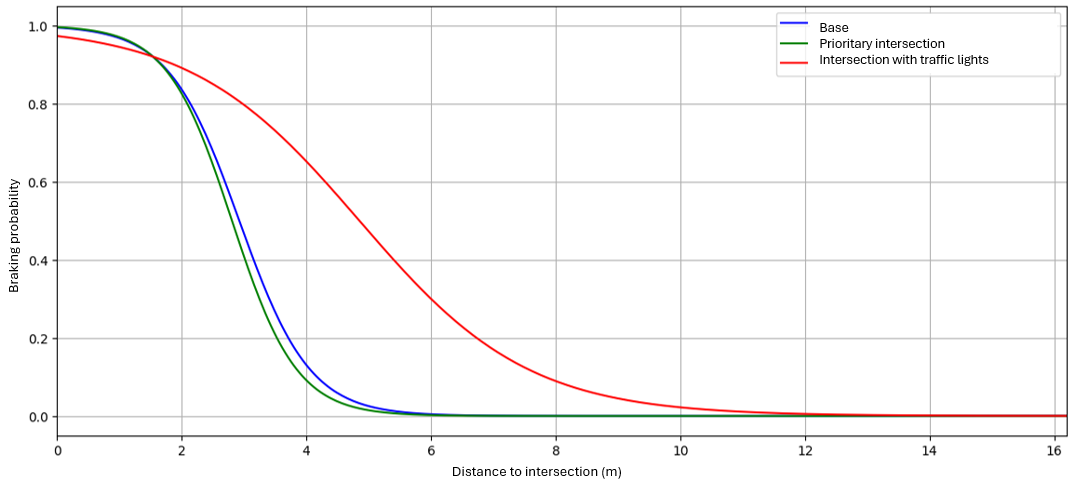}
    \caption{\textbf{Brake Probability vs. Distance to the Intersection.} Different types of intersections based on parameter estimates from the HM(IA) model.}
    \label{fig:braking}
\end{figure}

When analyzing intersection characteristics, it is evident that intersections with traffic lights significantly increase the probability of \textit{braking}, which is expected. A noteworthy point is that in the models that use data from the \textit{LVM}, the variable distinguishes whether the traffic light is red or not (“red light” variable). Models without LVM variables only consider whether the intersection has traffic lights or not. This change improves the significance of the variable, as it better captures the cyclists’ stopping behavior. Figure \ref{fig:braking} shows how different types of intersection affect braking distances according to the HM(IA) model. The effect of signalized intersections is particularly notable, as the model indicates that braking begins 5 meters earlier in these cases.

Regarding the number of vehicle lanes, an interesting observation is that a higher number of lanes increases the probability of stopping. This may be due to the relationship between larger streets and a greater presence of traffic lights or vehicle traffic that forces cyclists to stop. This behavior is significant and is observed across all models. For example, in the HM(IA) model, holding other variables constant, when the number of lanes is higher, the probability of \textit{waiting} is the highest at 28.50\%.

Among the latent variables, the effect of activation stands out, as it increases the probability of \textit{braking} and \textit{accelerating}. This effect, while small, is significant in the HM, HM(I), and HM(IA) models, suggesting that activation increases in situations where it is necessary to adjust speed, for instance, braking to avoid an obstacle or accelerating to change direction. Fatigue does not have a significant effect on actions in any of the models. One possible explanation is that fatigue already affects activation through the tonic component, and thus its effect on actions may already be accounted for. 

Analyzing the parameters of the HM(IA) model, it can be seen that high vehicular activity has a significant effect, increasing cyclists’ intentions to \textit{brake} (26.63\%), \textit{stop} (29.28\%), and/or \textit{decelerate} (16.49\%). Stressful situations (identified with ChatGPT) also have a significant effect on the likelihood of \textit{braking} and \textit{waiting}, making these actions more probable. Finally, poor infrastructure significantly increases the probability of waiting, suggesting that cyclists may be more susceptible to interruptions along segments with deficient infrastructure. 

Based on the analysis of the results of choice component, several key points stand out. Incorporating emotional state improves model prediction; in many cases, parameters became statistically significant only after including stress or fatigue in the analysis. The utility of the variables \textit{speed} and \textit{distance to intersection} is also noteworthy when describing the baseline behavior of cyclists; these variables and their respective levels account for approximately 13\% of the variance in the model data. Finally, the \textit{LVM} tool proves useful for obtaining contextual variables, not only to capture stress or fatigue, but also to identify variables that meaningfully explain cyclist behavior. This enables the discovery of new insights into cyclist behaviors and confirms certain hypotheses regarding their preferences under different infrastructure and traffic scenarios.

\begin{figure}[H]
    \centering
    \includegraphics[width=14cm]{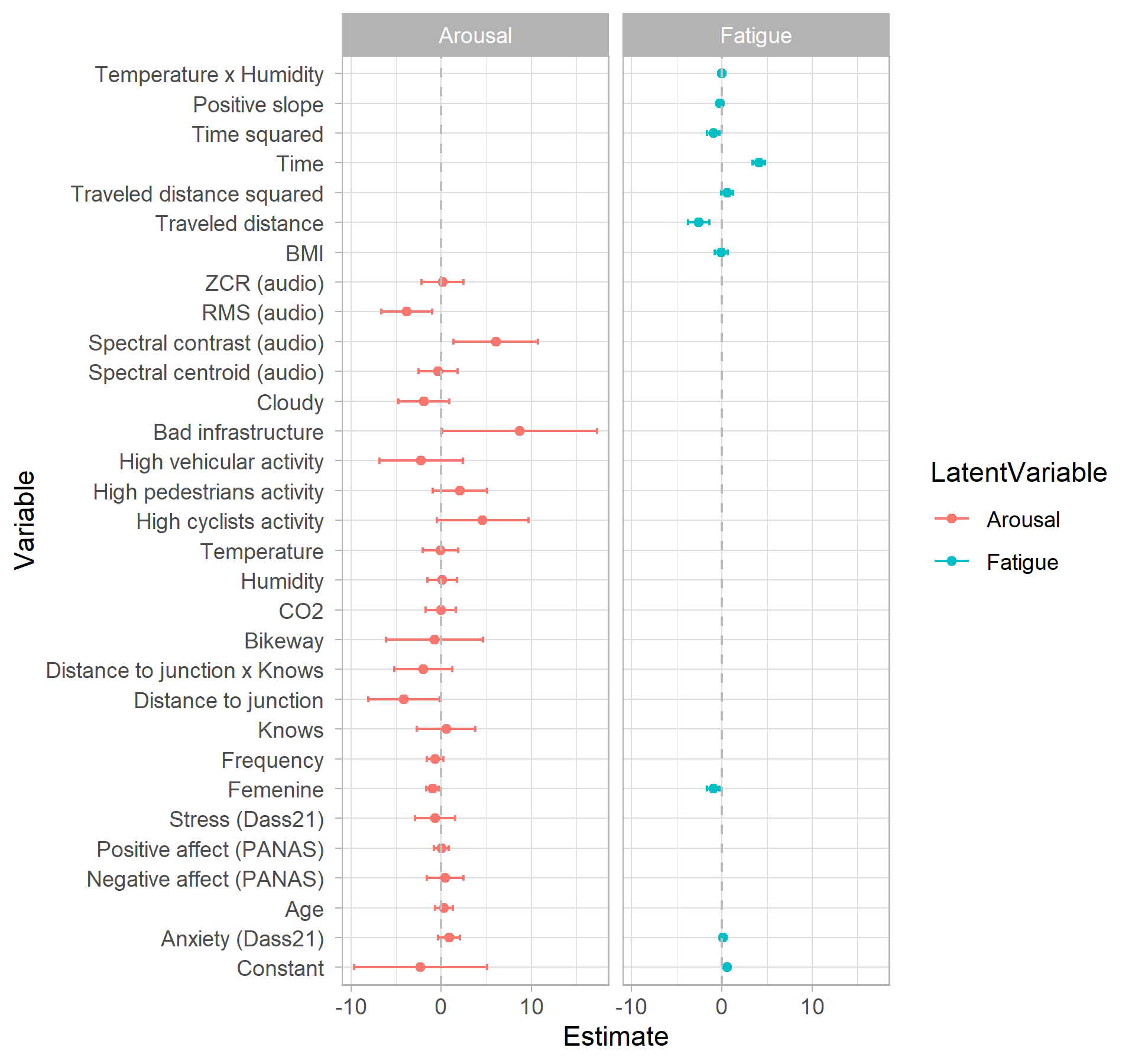}
    \caption{HM(IA) estimates and 90\% confidence interval, for latent arousal and fatigue.}
    \label{fig:output_latvar}
\end{figure}

Table \ref{tab:latent_comp}, in Appendix \ref{appendixtableresults}, shows the results of the latent component of the models, illustrating how different environmental variables affect both activation and fatigue. For analysis purposes, general comments are made for those referring specifically to the full model (HM(IA)). The results of arousal and fatigue latent variables in the HM(IA) model are summarized in \autoref{fig:output_latvar}.

Regarding arousal, it can be seen that there are not many significant variables overall; however, a few are significant and exhibit consistent behaviors across models. One such case is the distance to the intersection, which has a negative effect on arousal, meaning that as cyclists approach the intersection and the distance decreases, stress increases. This may be related to a higher perception of risk or the need to adjust behavior. 

Regarding contextual signals, such as $CO_2$, temperature and humidity, none have a significant effect on any of the models. An environmental variable that produces significant effects is \textit{cloudy} (obtained via \textit{LVM}), which appears to have a negative effect on activation, for example, in HM(I). This may be due to cloudy days that provide more pleasant conditions for physical activity.

Cycling frequency seems to have a negative effect on arousal, which means habitual cyclists become less stressed during the experiment. This is consistent with previous studies showing that experienced cyclists perceive less fear when cycling and are less likely to perceive a threat from motorized vehicles (\cite{rossetti2018modeling,rossetti2019want}). However, this effect is not consistent across all models. Most notably, the \textit{Bad infrastructure} variable (obtained through the LVM), shows a large increase in arousal. This confirms that poor infrastructure deteriorates the cycling experience, as widely reported in previous literature (\cite{heesch2012gender,wernerEvaluatingUrbanBicycle2019a,rossetti2018modeling,echiburu2021role}). Also noteworthy are the \textit{High pedestrians activity} and \textit{High cyclists activity} variables, which indicate that the presence of other pedestrians and cyclists increases cyclist stress. 

Regarding variables related to the baseline psychological state (Dass21 and PANAS), no relevant effects were found. Regarding demographic variables, only gender turned out to be significant, with females showing less arousal. Interestingly, \cite{henriquez2025identifying} found the contrary for public transportation trips, i.e. a larger arousal among female. Although this result may seem counterintuitive, the differences in cycling related to gender are complex and highly dependent on the context (\cite{prati2019gender, grudgings2018don, battiston2023revealing}), therefore, a lower overall arousal on women could be interpreted as either a positive or a negative perception, indicating the need for a deeper analysis of this variable. Regarding audio variables, all four studied signals were expected to have a positive effect on activation. One variable that shows this effect in a significant and consistent way is \textit{spectral contrast}. High \textit{spectral contrast} could indicate an environment with highly variable and contrasting sounds, such as traffic noise mixed with sirens, human voices, or machinery, creating a chaotic and potentially stressful soundscape. The other three variables do not show effects as strong as this one. Among them, \textit{RMS} stands out, which reaches a notable level of significance in the HM(IA) model and indicates that ambient noise level reduces activation. This result contradicts the initial hypothesis, and no clear explanations currently justify this behavior. 

Concerning fatigue, elapsed time is the variable that contributes the most. Traveled distance has a negative parameter, which seems counterintuitive but is consistent with findings of previous studies that indicate that, for relatively short trips, satisfaction with cycling commute can increase with distance due to the positive effect of exercise (\cite{olsson2013happiness, echiburu2021role}). The squared distance to the end of the trip and anxiety also have a non-negligible effect. One variable that also stands out is positive slope, which has a negative effect on fatigue. This can be explained by the design of the experimental route, where its first half (when the cyclists are less tired) involves an uphill segment, while its second half is downhill. To better capture this effect, the ideal setup would be a route with varying slopes at both the beginning and the end of the ride. On the other hand, it was found that females consistently exhibited lower fatigue levels. The body mass index (BMI) did not show a significant effect. A possible explanation for this behavior could again be its low variance due to a relatively young and healthy sample, with little variation in physical condition. The parameters and their significance levels show similar behavior across different versions of the proposed model, which is a good indicator of model robustness and the reliability of the estimated solutions.

Table \ref{tab:medition_comp} presents the measurement indicators for the latent variables of each model, along with their estimated parameters. This table is important for identifying whether an unobservable variable is being accurately measured. The shown parameters are consistent across the different models (MNL, HM, HM(I), HM(A), HM(IA)), indicating the robustness of the estimated relationships between observed and latent variables. Moreover, most parameters display high levels of statistical significance, suggesting that latent variables are reliably measured through the selected observed indicators. The only variable of concern is related to RMSSD, suggesting that the fatigue variable may not be adequately captured, indicating potential areas for improvement in the model. Regarding EDA's tonic component, it can be observed that fatigue has a significant positive effect on it ($\gamma_{F-PC}$), linked to increased sweating and electrodermal activity, consistent with findings in the literature. 

With respect to the measurement of the continuous component of choices, it is observed that higher speed is associated with a lower probability of accelerating; this can be seen in the parameter $\gamma_{Accelerate}$ in Table \ref{tab:medition_comp} across the different models. This result is in line with those reported by \textcite{SCHLEINITZ2017290}, who found that cyclists are less likely to accelerate after reaching their desired speed. Additionally, at higher speeds, braking becomes more likely than decelerating, although braking shows greater variability, suggesting that it occurs across a wider variety of situations.
\begin{table}[H]
	\centering
	\caption{Results of the latent variable measurement components (robust t-test in brackets).}
	\label{tab:medition_comp}
	\resizebox{\textwidth}{!}{%
		\begin{tabular}{llcccc}
			\hline
			\textbf{Component} & \textbf{Parameter} & HM & HM(I) & HM(A) & HM(IA) \\
			\hline			
			HR & $\gamma_{HR}$ & 0.18 (2.15**) & 0.22 (2.19**) & 0.21 (1.86*) & 0.21 (2.24**) \\
			& $\sigma_{HR}$ & 0.94 (20.84***) & 0.95 (20.90***) & 0.95 (20.81***) & 0.95 (20.87***) \\
			HRV & $\gamma_{HRV}$ & 0.01 (0.90) & 0.01 (0.93) & 0.01 (0.91) & 0.01 (0.94) \\
			 & $\sigma_{HRV}$ & 0.99 (239.21***) & 0.99 (239.14***) & 0.99 (238.94***) & 0.99 (238.92***) \\
			PC & $\gamma_{A-PC}$ & 0.00 (1.33) & 0.01 (1.74*) & 0.00 (1.75*) & 0.01 (1.77*) \\
			  & $\sigma_{PC}$ & 0.79 (12.49***) & 0.79 (12.53***) & 0.79 (12.57***) & 0.78 (12.54***) \\
			& $\gamma_{F-PC}$ & 0.32 (5.83***) & 0.39 (6.75***) & 0.38 (4.32***) & 0.38 (7.24***) \\
			TC & $\gamma_{TC}$ & 0.00 (1.59) & 0.01 (1.78*) & 0.01 (1.89*) & 0.02 (2.18**) \\
			& $\sigma_{TC}$ & 0.99 (228.71***) & 0.98 (197.83***) & 0.98 (170.34***) & 0.98 (159.41***) \\
			Accelerate & $\gamma_{Accelerate}$ & -0.47 (-8.02***) & -0.47 (-8.02***) & -0.47 (-8.02***) & -0.47 (-8.02***) \\
			Accelerate & $\sigma_{Accelerate}$ & 1.58 (18.85***) & 1.58 (18.85***) & 1.58 (18.85***) & 1.58 (18.85***) \\
			Decelerate & $\gamma_{Decelerate}$ & 0.78 (17.44***) & 0.78 (17.43***) & 0.78 (17.43***) & 0.78 (17.43***) \\
			Decelerate & $\sigma_{Decelerate}$ & 1.24 (33.40***) & 1.24 (33.41***) & 1.24 (33.41***) & 1.24 (33.41***) \\
			Brake & $\gamma_{Brake}$ & 1.29 (13.72***) & 1.29 (13.71***) & 1.29 (13.71***) & 1.29 (13.70***) \\
			Brake & $\sigma_{Brake}$ & 2.54 (41.32***) & 2.53 (41.36***) & 2.53 (41.36***) & 2.53 (41.37***) \\
			\hline
			\multicolumn{6}{r}{\normalsize\textbf{Note:} \normalsize{Stimated value (Robust t-test). * 90\%, ** 95\%, *** 99\%.}} \\	
			\hline
		\end{tabular}%
	}
\end{table}


\section{Discussion and conclusions}\label{sec:conclusion}

In this study, an ICLV model was proposed to analyze how different urban environmental variables and psychophysiological states, such as \textit{Activation} and \textit{Fatigue}, affect cyclist behavior in real urban environments. The aim of this study is to answer: P1. What urban environment and traffic characteristics affect cyclists' stress and fatigue? and P2. Are cycling decisions affected by latent psychological states of stress and fatigue? Despite previous laboratory evidence, to the extent of our knowledge, this is the first study providing field evidence to answer these questions.

The results confirm that cycling decisions are influenced by arousing (stress-related) emotions and demonstrate the significant impact of urban characteristics and traffic conditions on cyclist behavior. Factors such as distance to intersections, cyclist speed, number of lanes, vehicular activity, and infrastructure quality play key roles in modulating both psychological arousal and cycling decisions. Notably, some qualitative variables—such as vehicular activity and infrastructure quality—were incorporated into the model solely based on video interpretations generated by a large language model (LLM).

One of the most notable aspects of the proposed model is the inclusion of physiological indicators as latent variables, which enables capturing the complexity of cyclists' emotional and physical responses to different environmental stimuli. Throughout the modeling process, it has been shown that incorporating these latent variables significantly improves the predictive capacity of the model, as evidenced by the fit metrics when compared to classic discrete choice models (MNL). This improvement is particularly notable when incorporating contextual data derived from real ride images, such as the image analysis enabled by an LLM like the proposed \textit{LVD} method, which has made it possible to obtain contextual variables with an unprecedented level of detail and personalization.

Innovation in methods for obtaining contextual variables—particularly the use of language models to generate image-based descriptions—represents a significant advancement in urban mobility research. These methods overcome the limitations of traditional approaches by providing rich and customizable descriptions of the environment, tailored to the specifics of each study. Furthermore, their ability to capture detailed elements of the surroundings, such as infrastructure quality or vehicular and pedestrian activity, has been essential in identifying the key factors influencing cyclists’ activation and fatigue.

In terms of recommendations derived from this study for cyclist infrastructure design, several points stand out: feeling safe during their journeys is crucial for cyclists, especially for those with less experience. Considering the variables that most significantly affect activation, it is recommended to design cyclist-friendly intersections. Cyclists clearly prefer bike lanes over other types of roads with pedestrians or vehicles. The presence of other cyclists appears to be a source of stress, which supports the case for creating wide bike lanes where the risk of collisions is lower. Interaction with pedestrians also contributes to discomfort, suggesting that the ideal bike lane should be segregated from both motor vehicles and pedestrians. Another important aspect is to ensure infrastructure allows cyclists to maintain their desired speed without frequent interruptions.

The main limitation of this study is the absence of direct information about the cyclist’s actual decisions, meaning that imputation relies on observed changes in speed. We can just infer their decisions based on the observed behavior. This leads to a classification in which braking and decelerating are mainly distinguished by the magnitude and abruptness of the speed change. Specifically, braking is associated with a sharp decrease, while decelerating may represent micro-braking or simply reduced pedaling. Another limitation is caused by the homogeneity of the sample. Participants were young students, with relatively good health conditions, similar levels of physical activity, and similar base stress levels. 

In the estimated models, the effect of audio was not as significant as that of images. Given that the results show a limited impact for 3 of the 4 proposed audio variables and considering the relatively scarce existing evidence regarding the influence of such data in the specific context of this study, further research is recommended to validate their usefulness, or alternatively, to use only spectral contrast, which was the signal that provided a meaningful interpretation. 

Future research should consider expanding the use of LLM for urban analysis. The capabilities of LLM for interpreting urban and traffic conditions allow the systematization of infrastructure qualitative analysis and obtaining inputs for user experience analysis. Aerial imagery could be used to expand this methodology to a broader area and to detect zones that may be stressful or risky for different users and travel modes.

In conclusion, the results of this study not only confirm the importance of latent variables in modeling cyclist behavior, but also highlight the value of a multimodal approach to obtaining contextual data. The combination of psychophysiological indicators with innovative methods for contextual analysis has enabled a more precise and detailed characterization of the urban environment, which, in turn, has enhanced the model’s ability to predict and explain cyclists' behavior. This multidimensional approach not only contributes to a better understanding of urban mobility, but also provides valuable tools for designing policies and strategies that promote safer and more efficient mobility in cities.

\vspace{10pt}

\noindent \textbf{ACKNOWLEDGEMENTS}

\noindent This research was partially funded by ANID, FONDECYT 1231584, ANID PIA/PUENTE AFB220003 and ANID, FONDEF IT21I0059. Analysis was developed with open-source software R (\cite{R2023}) using the supercomputing infrastructure of the NLHPC (CCSS210001).

\printbibliography

\newpage
\appendix	

	\setcounter{table}{0}
    \renewcommand{\thetable}{A\arabic{table}}
    \renewcommand*{\theHtable}{\thetable}

\section{Action Imputation Algorithm}
\label{appendix:algoritmo_acciones}

\noindent The action assignment algorithm developed for this study aims to identify and classify the predominant actions of a cyclist within 5-second time windows, using as input instantaneous speed data recorded at a frequency of one second. The main limitation of this approach is the absence of direct information on the cyclist’s actual actions, so imputation is based on observed changes in speed. This results in a classification where braking and decelerating are mainly distinguished by the magnitude and abruptness of the speed change. Specifically, braking is associated with a sharp decrease, while decelerating may reflect micro-braking or simply reduced pedaling.

\noindent\textbf{Algorithm Description}\\
The process begins with the initial classification of actions based on the difference in speed between consecutive records. Thresholds are defined to determine whether a decrease in speed corresponds to braking or deceleration, as well as to identify acceleration. The steps of the algorithm are as follows:

\begin{enumerate}
	\item \textbf{Speed Difference Calculation}: The difference in speed between consecutive rows in the DataFrame is calculated. This difference serves as the basis for identifying changes in cyclist behavior.
	\item \textbf{Initial Action Classification}: An initial action is defined for each row of data based on the magnitude of the speed change. Possible actions include:
	\begin{itemize}
		\item \textit{Brake}: Identified when the decrease in speed is greater than or equal to the braking threshold.
		\item \textit{Decelerate}: Corresponds to a decrease in speed smaller than the braking threshold but greater than the deceleration threshold.
		\item \textit{Accelerate}: Assigned when there is an increase in speed greater than or equal to the acceleration threshold.
		\item \textit{Wait}: Assigned when the speed is close to zero.
		\item \textit{Maintain Speed}: Assigned when no significant changes in speed are detected.
	\end{itemize}
	\item \textbf{Grouping Actions into Time Windows}: The classified actions are grouped into 5-second time windows. In each window, the predominant action is selected. To resolve conflicts, actions are prioritized in the following order: brake, wait, accelerate, decelerate, and maintain speed.
	\item \textbf{Assignment Correction}: After the initial grouping, various functions are applied to review and correct the action assignments.
\end{enumerate}

\noindent Finally, the algorithm returns a DataFrame with the predominant actions assigned to each time window, along with the accumulated magnitudes of acceleration, deceleration, and braking in each window.\\

This approach, although based on limited data, provides a reasonable approximation for identifying cyclist actions, offering valuable information for behavioral analysis across different segments of the route.

	\section{LLM Video descriptor (LVD) - Prompt}\label{anexoA}
    

\label{appendix:gpt_prompt}

\begin{small}
\begin{verbatim}
Instructions for Describing Bicycle Image Sequences:
For each image sequence, provide a detailed description based on the following
characteristics:

LaneType: Single word describing if the bicycle lane is:
"dedicated"
"shared"
"does not exist"

Type of bicycle lane separation: Description of the type of separation between the bike lane
and the rest of the road:

"Physical"
"visual"
"none"

TrafficSignal: Traffic signal facing the cyclist:
"Green"
"Red"
"Not identifiable" if no clear color is seen on the traffic light
"Not present" if there is no traffic light in the images.

Signage: Concise list of observed traffic signs:
Examples: "bike route", "yield", "stop", "no parking", "no turn", "crossing warning",
"none".

VehicleProximity: Proximity of moving vehicles:
Car: "High", "Medium", "Low", "Not present"
Truck: "High", "Medium", "Low", "Not present"
Motorcycle: "High", "Medium", "Low", "Not present"
Bicycle: "High", "Medium", "Low", "Not present"

Type of Nearby Pedestrian: Proximity of pedestrians:
Adult: "High", "Medium", "Low", "Not present"
Child: "High", "Medium", "Low", "Not present"
Group: "High", "Medium", "Low", "Not present"
Pet: "High", "Medium", "Low", "Not present"

Road Condition: General state of the pavement:
"good"
"fair"
"poor"

Presence of potholes: Indication of potholes on the road:
"present"
"not present"

Pedestrian Activity: General proximity of pedestrians:
"High"
"Medium"
"Low"

Obstructions: Presence of obstructions in the middle of the path:
"Present"
"Not present"

WeatherCondition: Weather condition during the sequence:
"Sunny"
"Cloudy"

CyclistStressLevel: Subjective stress level of the cyclist based on the entire environment:
"High"
"Medium"
"Low"

StressLevelDescription: Justified description explaining why the CyclistStressLevel was assigned.

Special Events: Record of any special event that may affect the cyclist's stress or require them
to react in some way:
"Present"
"Not present"
Examples: a vehicle or person passing very close, an obstacle on the route, a vehicle that
crosses in the middle of the path, etc.

Road Works: "Present", "Not present"

Presence of Other Cyclists: "High", "Medium", "Low", "Not present"

Cyclist Infrastructure: Quality of the bike lane:
"good"
"fair"
"poor"

Considerations: The images were obtained from a fixed route within Santiago de Chile; some
bike lanes may not look as usual. In many cases, when smaller areas separated with some type
of lane delineator are seen, it means there is a bike lane present.
All images are captured from a GoPro mounted on the bicycle's handlebar. Since the
cyclist tends to tilt the handlebar when stopping, some images may appear tilted
from the street's perspective. Each sequence is approximately 5 seconds long.
\end{verbatim}
\end{small}

\section{LLM Video Descriptor (LVD) - Results} \label{appendix:gpt_results}

\begin{figure}[H]
\centering
\begin{subfigure}{0.44\textwidth}
    \includegraphics[width=\textwidth]{img/bad_infra.png}
    \caption{Poor Cycling Infrastructure.}
    \label{fig:bad_infra}
\end{subfigure}
\hfill
\begin{subfigure}{0.44\textwidth}
    \includegraphics[width=\textwidth]{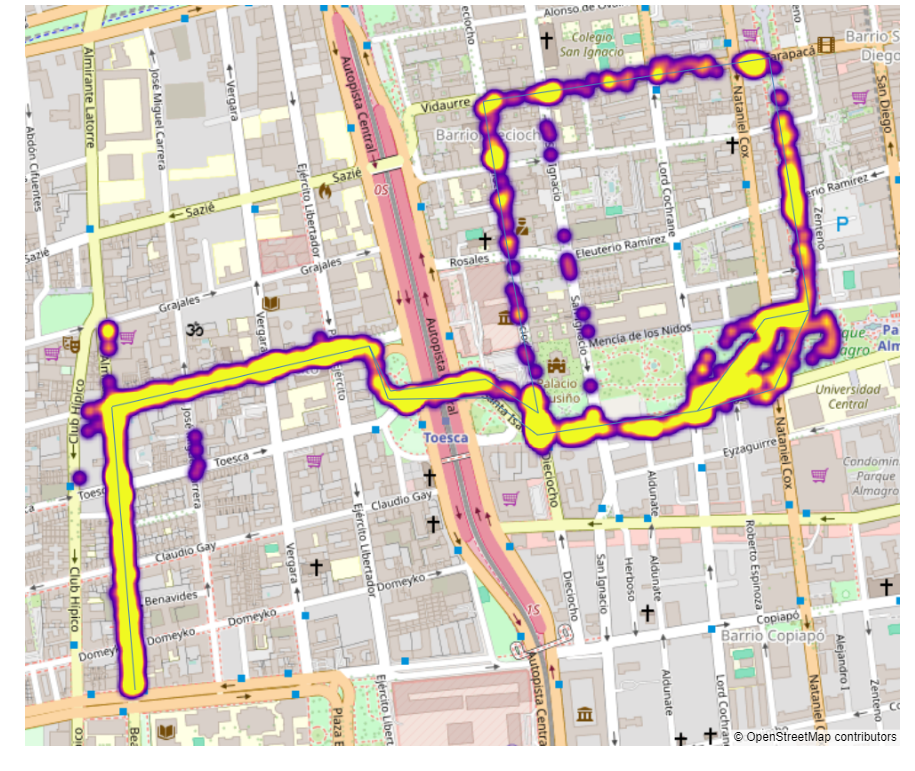}
    \caption{Poor Road Condition.}
    \label{fig:bad_road_cond}
\end{subfigure}

\vspace{0.3cm}

\begin{subfigure}{0.44\textwidth}
    \includegraphics[width=\textwidth]{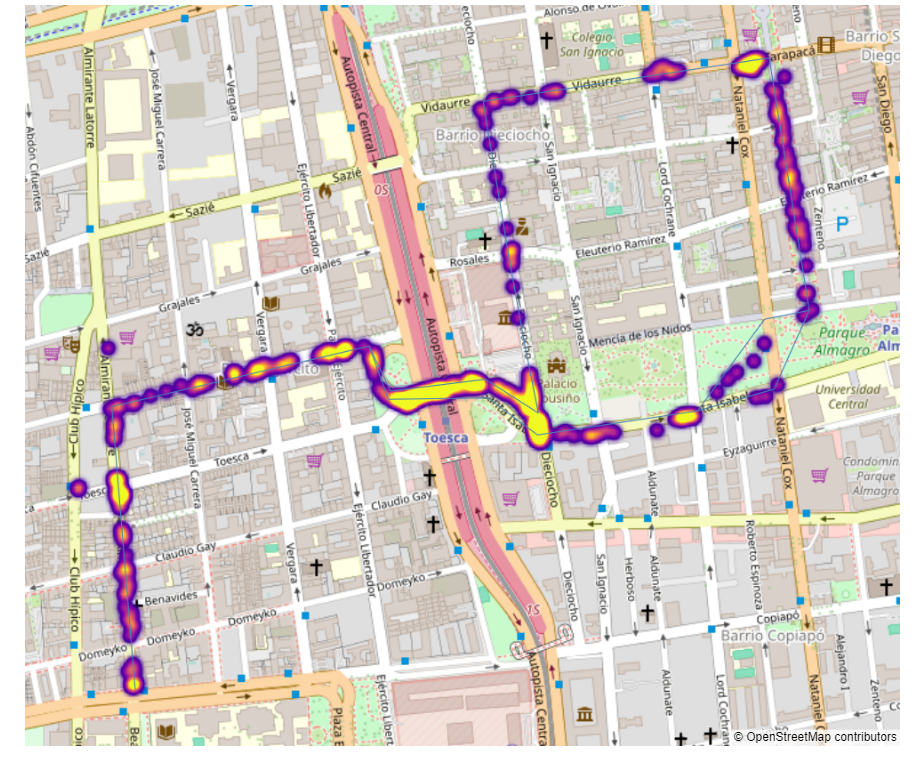}
    \caption{High Cyclist Proximity.}
    \label{fig:high_bikers_level}
\end{subfigure}
\hfill
\begin{subfigure}{0.44\textwidth}
    \includegraphics[width=\textwidth]{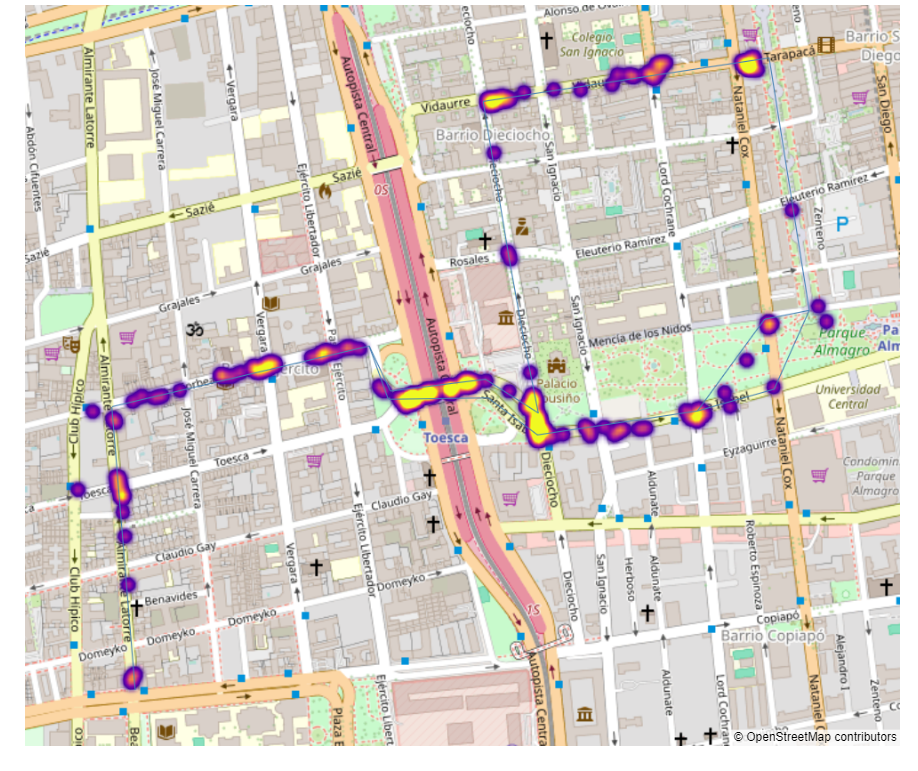}
    \caption{High Motorcycle Proximity.}
    \label{fig:high_motorbikes_prox}
\end{subfigure}

\vspace{0.3cm}

\begin{subfigure}{0.44\textwidth}
    \includegraphics[width=\textwidth]{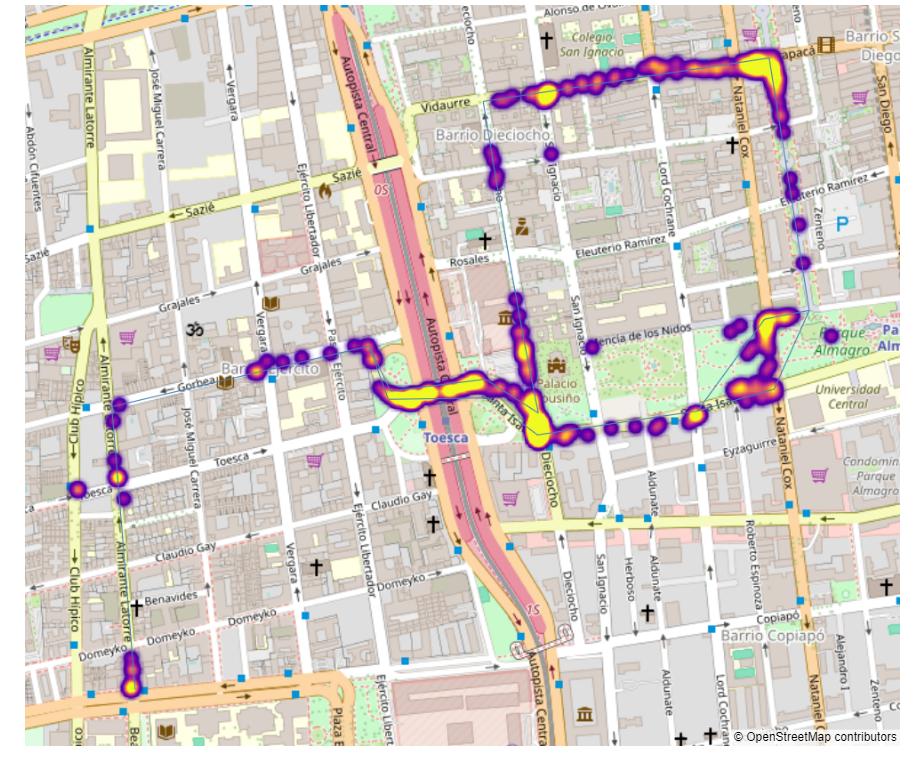}
    \caption{High Bus Proximity.}
    \label{fig:high_buses_act}
\end{subfigure}
\hfill
\begin{subfigure}{0.44\textwidth}
    \includegraphics[width=\textwidth]{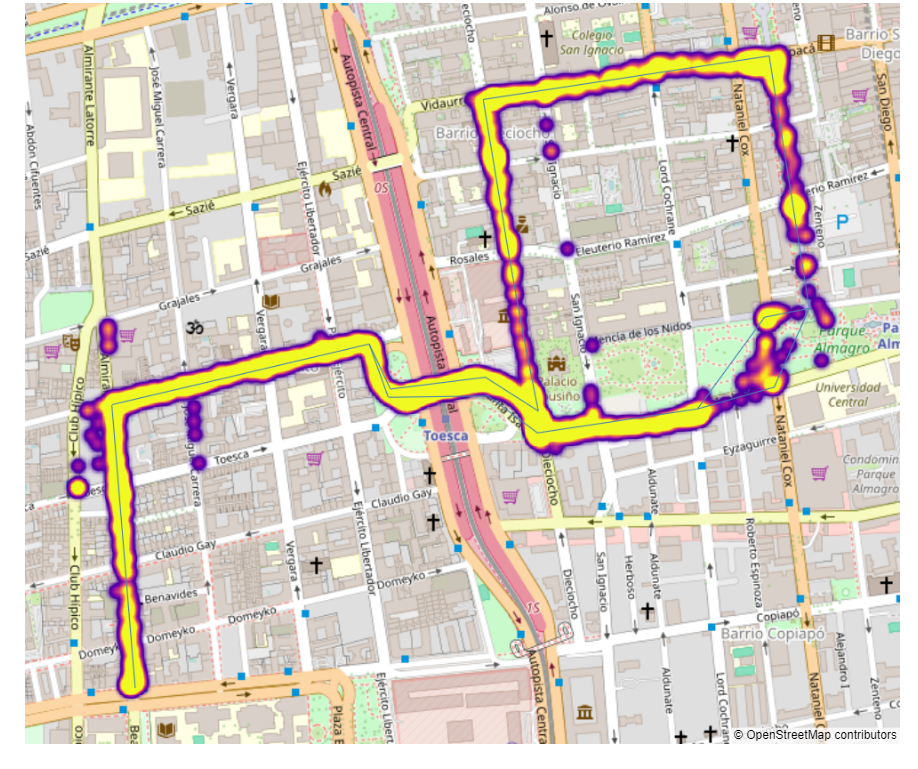}
    \caption{High Car Proximity.}
    \label{fig:high_cars_prox}
\end{subfigure}

\caption{Heatmaps generated by LVD (Part 1). One participant briefly deviated from the designated route, which is reflected in the spatial distribution of the data—noticeable in Figures A and B.}
\label{fig:gpt_image_descriptor_results_part1}
\end{figure}

\begin{figure}[H]
\centering
\begin{subfigure}{0.44\textwidth}
    \includegraphics[width=\textwidth]{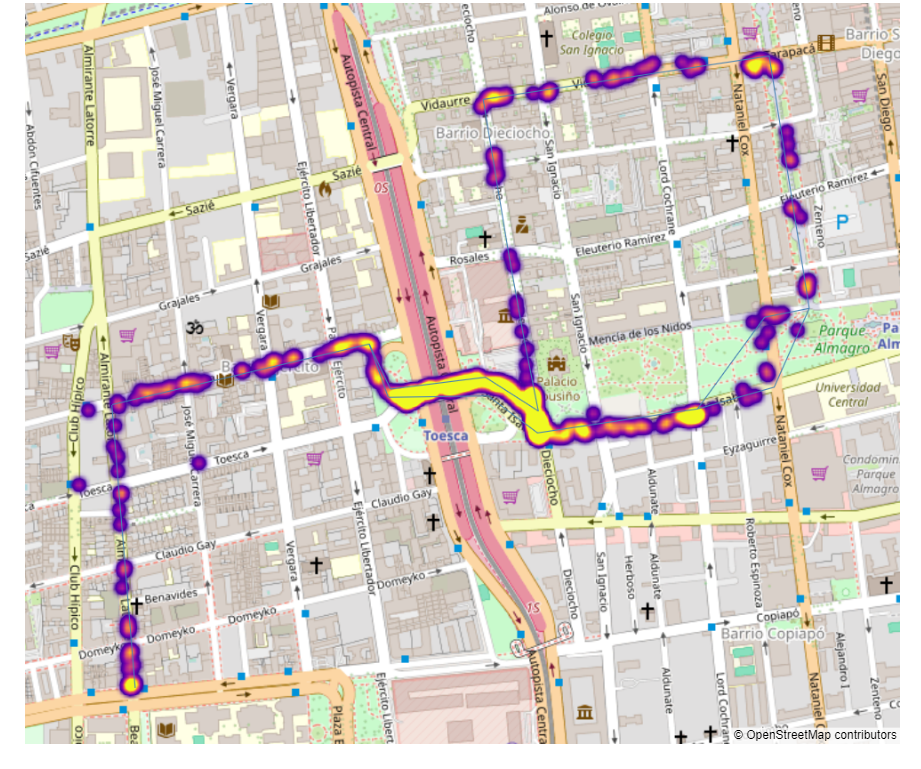}
    \caption{High Truck Proximity.}
    \label{fig:high_trucks_act}
\end{subfigure}
\hfill
\begin{subfigure}{0.44\textwidth}
    \includegraphics[width=\textwidth]{img/red_traffic_signal.png}
    \caption{Red Traffic Lights.}
    \label{fig:red_traffic_signal}
\end{subfigure}

\vspace{0.3cm}

\begin{subfigure}{0.44\textwidth}
    \includegraphics[width=\textwidth]{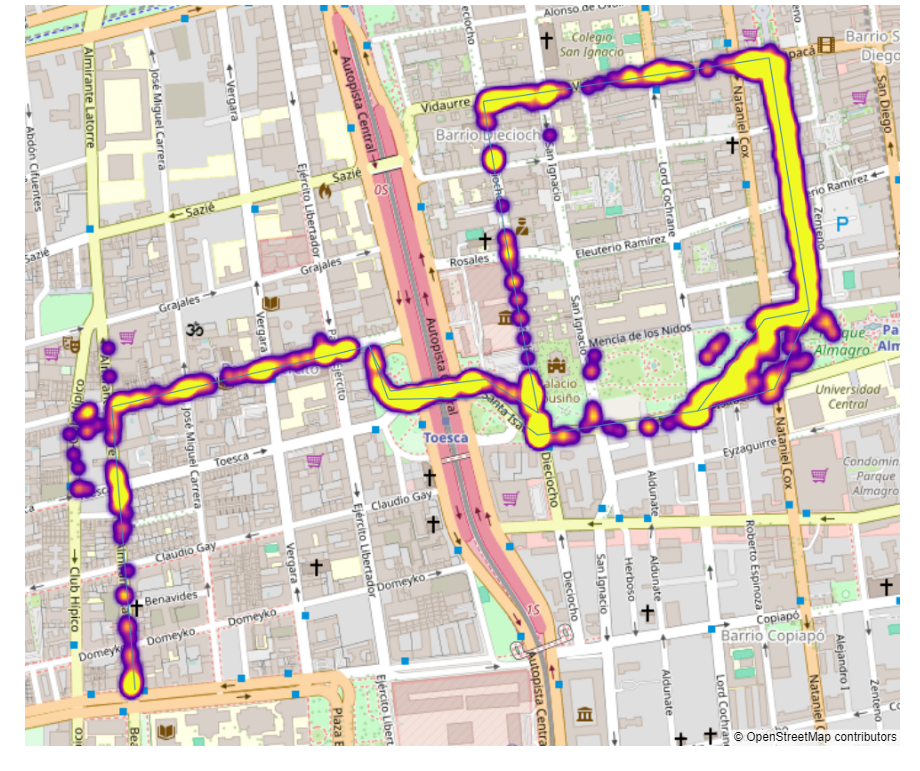}
    \caption{High Pedestrian Proximity.}
    \label{fig:high_pedestrians_act}
\end{subfigure}
\hfill
\begin{subfigure}{0.44\textwidth}
    \includegraphics[width=\textwidth]{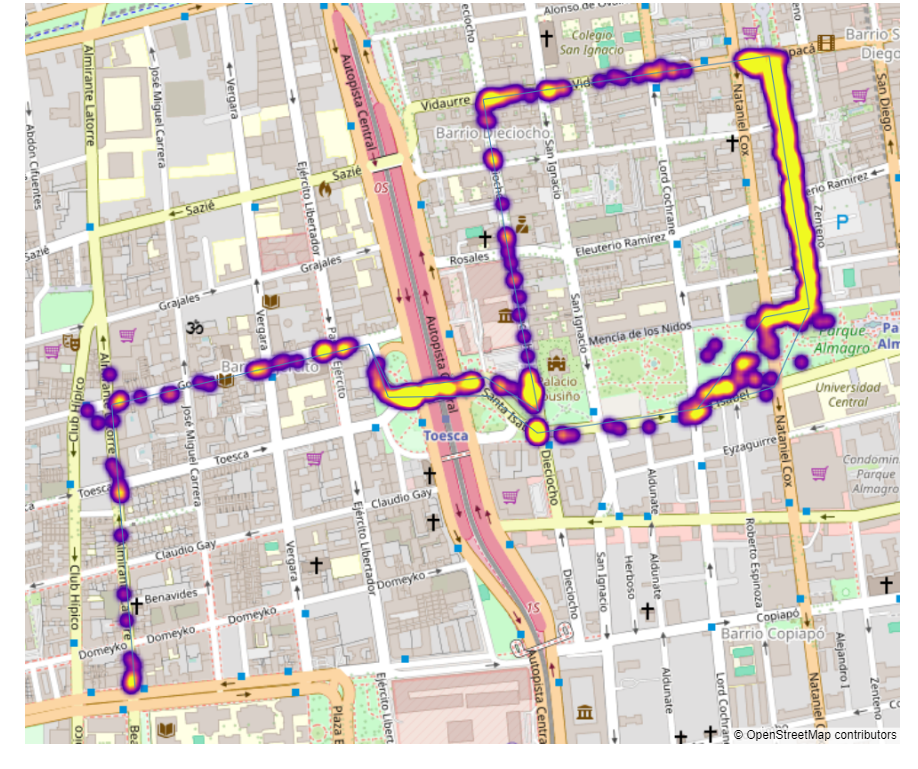}
    \caption{High Group Proximity (Pedestrians).}
    \label{fig:high_groups_act}
\end{subfigure}

\vspace{0.3cm}

\begin{subfigure}{0.44\textwidth}
    \includegraphics[width=\textwidth]{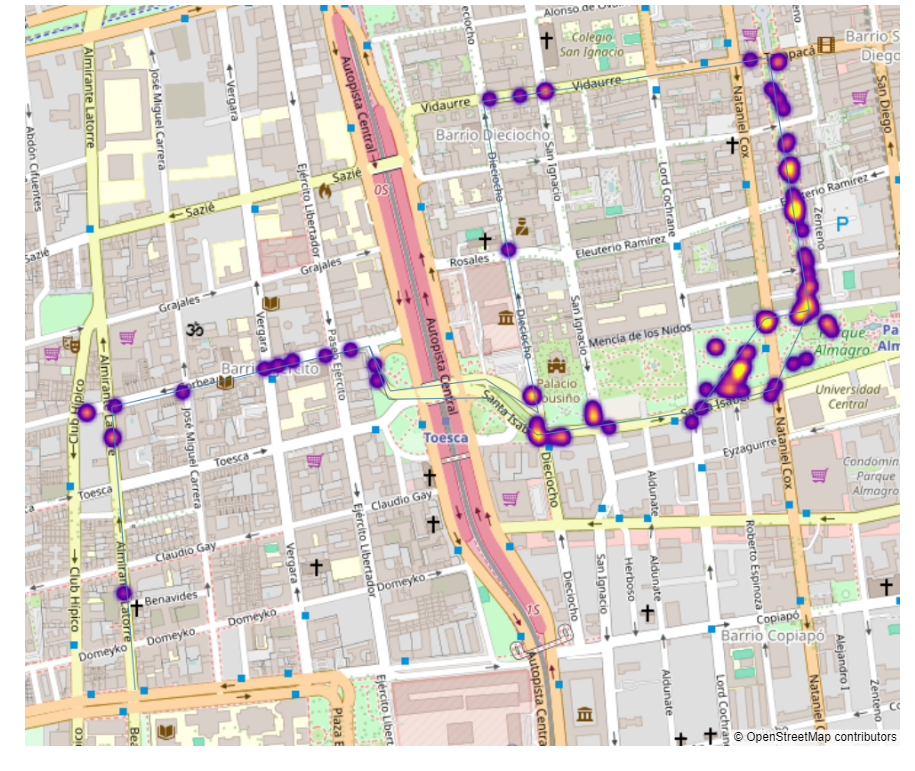}
    \caption{High Proximity to Children or Pets.}
    \label{fig:high_kids_act}
\end{subfigure}
\hfill
\begin{subfigure}{0.44\textwidth}
    \includegraphics[width=\textwidth]{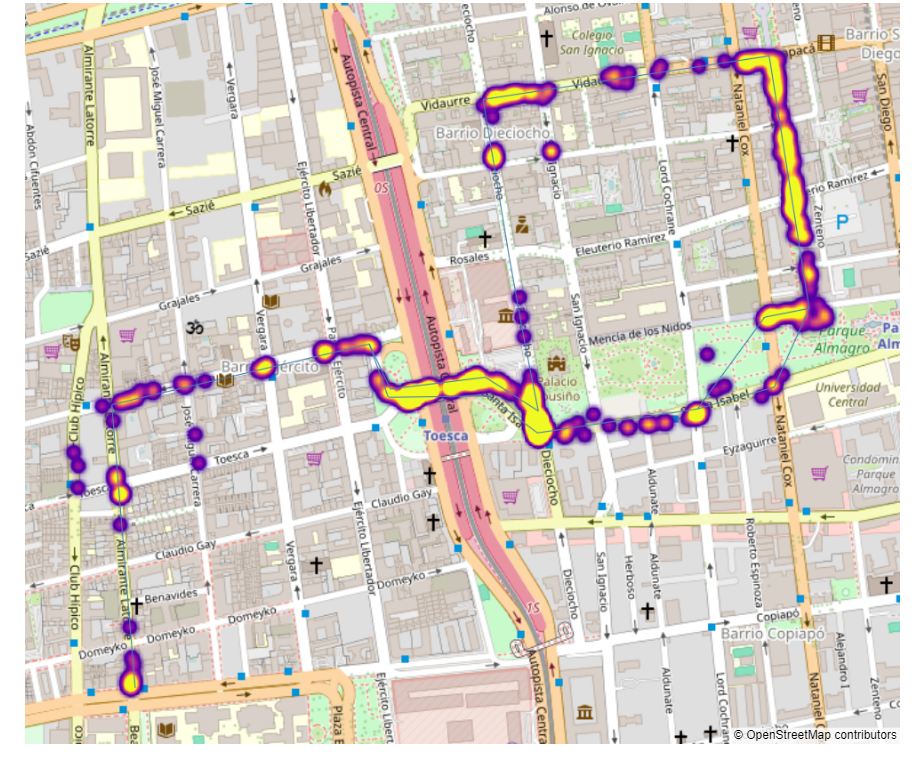}
    \caption{Stressful Situations.}
    \label{fig:highstress}
\end{subfigure}

\caption{Heatmaps generated by LVD (Part 2).}
\label{fig:gpt_image_descriptor_results_part2}
\end{figure}

\section{Results tables}\label{appendixtableresults}

\begin{table}[H]
\centering
\caption{Results of the choice component of the models (robust standard errors and robust t-ratio in brackets).}
\label{tab:choice_component}
\resizebox{\textwidth}{!}{%
\begin{tabular}{llcccccc}
\hline
\textbf{Variable} & \textbf{Source} & \textbf{MNL} & \textbf{MNL(I)} & \textbf{HM} & \textbf{HM(I)} & \textbf{HM(A)} & \textbf{HM(IA)} \\
\hline
\textbf{Accelerate} \\[2pt]
Constant & Constant & 0.84 (4.48***) & 0.58 (3.29***) & 1.24 (1.88*) & 1.74 (1.79*) & 0.98 (5.00***) & 0.58 (2.89***) \\[2pt]

Arousal & Latent &            &            & 0.44 (4.45***) & 0.37 (9.05***) & -0.01 (-1.42)   & 0.13 (-1.85**) \\
Fatigue & Latent &            &            & 0.03 (1.34)    & 0.04 (1.29)    & 0.04 (1.44)     & 0.03 (1.13)     \\[2pt]

Distance to junction & Distance to junction & -1.32 (-7.56***) & -1.21 (-6.73***) & 8.35 (1.49)  & 1.57 (0.89)   & -1.38 (-7.14***) & -1.29 (-6.47***) \\
High distance to junction & Distance to junction & 0.88 (6.33***) & 0.84 (5.97***) & 0.91 (6.40***) & 0.86 (6.05***) & 0.85 (5.89***)   & 0.81 (5.74***)   \\
Low distance to junction & Distance to junction & 1.44 (8.38***) & 1.31 (7.43***) & 1.46 (7.98***) & 1.31 (7.05***) & 1.42 (8.13***)   & 1.30 (7.43***)   \\
Distance to junction x Knows & Distance to junction & -0.08 (-0.74) & -0.08 (-0.73) & -0.01 (-0.08)  & -0.00 (-0.03)  & -0.16 (-1.18)    & -0.16 (-1.21)    \\[2pt]

Speed & Speed & -1.03 (-6.73***) & -1.07 (-6.91***) & -1.11 (-7.72***) & -1.16 (-8.33***) & -1.11 (-6.84***) & -1.15 (-7.13***) \\
High speed & Speed & 0.47 (2.78***) & 0.49 (2.90***) & 0.31 (2.07**) & 0.32 (2.09**) & 0.44 (2.55**) & 0.46 (2.68***)  \\
Low speed & Speed & 1.90 (10.65***) & 1.84 (10.32***) & 2.12 (12.07***) & 2.07 (12.16***) & 1.90 (10.67***) & 1.84 (10.43**)  \\[2pt]

Junction & Infrastructure & 1.04 (6.64***) & 0.89 (6.77***) & 1.06 (6.00***) & 0.89 (5.92***) & 1.00 (6.09***) & 0.84 (5.99***) \\
N° car lanes & Infrastructure & 0.12 (2.10**) & 0.17 (2.72***) & 0.21 (3.84***) & 0.27 (3.87***) & 0.04 (0.67)   & 0.15 (2.13**)   \\[2pt]

High vehicular activity & GPT &             & 0.20 (2.06**) &             & 1.46 (1.11)   &             & 0.10 (0.64)    \\
Bad infrastructure & GPT &             & 0.22 (1.95*) &             & -5.83 (-1.69*) &             & 0.49 (2.06**)   \\[6pt]

\textbf{Decelerate} \\[2pt]
Constant & Constant & 0.59 (3.05***) & 0.01 (0.05) & 0.83 (1.71*) & 0.63 (1.11) & 0.53 (2.58***) & 0.00 (0.01) \\[2pt]

Arousal & Latent &            &            & 0.25 (2.53**) & 0.20 (2.95***) & 0.00 (1.01) & 0.01 (0.62) \\
Fatigue & Latent &            &            & -0.00 (-0.04) &               & -0.00 (-0.12) & -0.00 (-0.07) \\[2pt]

Distance to junction & Distance to junction & -1.62 (-7.61***) & -1.48 (-7.34***) & 3.95 (0.87) & 0.02 (0.02) & -1.58 (-7.26***) & -1.45 (-6.80***) \\
High distance to junction & Distance to junction & 0.71 (4.40***) & 0.65 (4.12***) & 0.72 (4.39***) & 0.66 (4.03***) & 0.71 (4.49***) & 0.65 (4.10***) \\
Low distance to junction & Distance to junction & 1.90 (7.31***) & 1.74 (7.10***) & 1.91 (7.14***) & 1.74 (6.87***) & 1.92 (7.49***) & 1.74 (7.15***) \\[2pt]

Speed & Speed & 0.07 (0.47)  & 0.08 (0.53)  & 0.01 (0.06)  & 0.02 (0.13)  & 0.10 (0.65)  & 0.10 (0.59)  \\
High speed & Speed & 0.25 (1.59) & 0.25 (1.59) & 0.17 (1.16) & 0.17 (1.15) & 0.25 (1.52) & 0.25 (1.54) \\
Low speed & Speed & 1.64 (6.97***) & 1.58 (6.67***) & 1.78 (7.13***) & 1.71 (6.84***) & 1.65 (7.04***) & 1.59 (6.76***) \\[2pt]

Junction & Infrastructure & 0.29 (1.20) & 0.12 (0.51) & 0.30 (1.35) & 0.13 (0.60) & 0.31 (1.29) & 0.12 (0.52) \\
N° car lanes & Infrastructure & -0.01 (-0.27) & 0.11 (2.10**) & 0.04 (0.88) & 0.17 (3.00***) & 0.02 (0.38) & 0.12 (2.16**) \\[2pt]

High vehicular activity & GPT &            & 0.28 (3.48***) &            & 0.95 (1.31)   &            & 0.30 (3.97***)  \\
Bad infrastructure & GPT &            & 0.51 (5.49***) &            & -2.67 (-1.26) &            & 0.47 (3.84***)  \\
Route in bad condition & GPT &            & 0.22 (2.77***) &            & 0.21 (2.64***) &            & 0.21 (2.68***)  \\[6pt]

\textbf{Wait} \\[2pt]
Constant & Constant & -0.64 (-2.78***) & -1.99 (-7.76***) & -0.69 (-2.97***) & -2.06 (-7.91***) & -0.64 (-2.84***) & -1.99 (-7.76***) \\[2pt]


Distance to junction & Distance to junction & -3.77 (-9.03***) & -3.65 (-9.51***) & -3.78 (-8.96***) & -3.64 (-9.44***) & -3.77 (-9.01***) & -3.65 (-9.48***) \\
High distance to junction & Distance to junction & 2.29 (4.53***) & 2.35 (4.72***) & 2.30 (4.53***) & 2.34 (4.68***) & 2.29 (4.52***) & 2.35 (4.70***) \\
Low distance to junction & Distance to junction & 4.07 (10.05***) & 4.06 (11.19***) & 4.08 (9.98***) & 4.06 (11.17***) & 4.07 (10.03***) & 4.05 (11.14***) \\[2pt]


N° car lanes & Infrastructure & 0.37 (3.82***) & 0.60 (5.90***) & 0.40 (3.92***) & 0.63 (5.94***) & 0.37 (3.88***) & 0.60 (5.92***) \\[2pt]

High vehicular activity & GPT &            & 0.86 (7.19***) &            & 0.87 (7.10***) &            & 0.86 (7.18***) \\
Bad infrastructure & GPT &            & 1.28 (8.15***) &            & 1.30 (8.27***) &            & 1.28 (8.11***) \\
Route in bad condition & GPT &            & 0.25 (1.96**)  &            & 0.27 (2.03**)  &            & 0.24 (1.93*)   \\
Stressful situation & GPT &            & 0.48 (3.26***) &            & 0.47 (3.00***) &            & 0.48 (3.21***) \\[6pt]

\textbf{Brake} \\[2pt]
Constant & Constant & 0.65 (3.03***) & -0.24 (-1.08) & 1.12 (1.31) & 1.20 (0.98) & 0.38 (1.42) & -0.31 (-1.16) \\[2pt]

Arousal & Latent &            &            & 0.57 (5.57***) & 0.47 (8.67***) & 0.02 (1.54) & 0.05 (1.71*) \\
Fatigue & Latent &            &            & 0.03 (0.84)    & 0.02 (0.49)    & 0.02 (0.49) & 0.02 (0.51)   \\[2pt]

Distance to junction & Distance to junction & -3.00 (-14.05***) & -2.94 (-14.10***) & 9.60 (1.35)  & 0.65 (0.30)   & -2.84 (-12.83***) & -2.72 (-11.85***) \\
High distance to junction & Distance to junction & 1.27 (7.03***) & 1.33 (7.20***) & 1.30 (7.24***) & 1.35 (7.27***) & 1.32 (7.28***) & 1.35 (7.00***) \\
Low distance to junction & Distance to junction & 3.81 (12.67***) & 3.77 (12.92***) & 3.86 (12.42***) & 3.78 (12.39***) & 3.81 (12.38***) & 3.75 (12.40***) \\[2pt]

Speed & Speed & -0.77 (-4.78***) & -0.65 (-3.89***) & -0.88 (-5.93***) & -0.77 (-5.14***) & -0.61 (-3.56***) & -0.50 (-2.84***) \\
High speed & Speed & 0.89 (5.08***) & 0.84 (4.67***) & 0.70 (4.05***) & 0.65 (3.71***) & 0.87 (4.98***) & 0.83 (4.49***) \\
Low speed & Speed & 2.83 (12.73***) & 2.79 (12.60***) & 3.11 (15.47***) & 3.07 (15.16***) & 2.86 (12.82***) & 2.82 (12.37**) \\[2pt]

Yield or stop sign & Infrastructure & 0.33 (0.75)  & -0.33 (-1.61) & 0.38 (0.87)  & -0.31 (-1.52) & 0.43 (1.00)   & -0.38 (-1.72*) \\
Junction & Infrastructure& -0.38 (-0.71) & 0.37 (1.02) & -0.34 (-0.63) & 0.43 (1.11) & -0.60 (-1.05) & 0.31 (0.79) \\
N° car lanes & Infrastructure & 0.06 (0.91) & 0.23 (3.12***) & 0.19 (2.60***) & 0.35 (3.97***) & 0.21 (2.12**) & 0.29 (2.95***) \\
Traffic ligth & Infrastructure & 1.48 (3.01***) &             & 1.52 (3.06***) &             & 1.63 (3.29***) &             \\[2pt]

High vehicular activity & GPT &            & 0.76 (7.86***) &            & 2.36 (1.43)   &            & 0.94 (5.76***) \\
Bad infrastructure & GPT &            & 0.77 (5.37***) &            & -6.89 (-1.62) &            & 0.27 (0.86)    \\
Red traffic ligth & GPT &            & 1.68 (4.34***) &            & 1.70 (4.41***) &            & 1.60 (4.08***) \\
Stressful situation & GPT &            & 0.56 (3.63***) &            & 0.51 (3.44***) &            & 0.52 (3.47***) \\
\hline
\multicolumn{8}{r}{\normalsize\textbf{Note:} \normalsize{Estimated value (Robust t-test). * 90\%, ** 95\%, *** 99\%.}} \\
\hline
\end{tabular}%
}
\end{table}

\begin{table}[H]
\centering
\caption{Results of the latent variable component of the models (robust t-test in brackets).}
\label{tab:latent_comp}
\resizebox{\textwidth}{!}{%
\begin{tabular}{llcccc}
\hline
\textbf{Parameter} & \textbf{Source} & \textbf{HM} & \textbf{HM(I)} & \textbf{HM(A)} & \textbf{HM(IA)} \\
\hline
\textbf{Arousal} \\[3pt]

Constant & Constant & -0.49 (-0.34)    & -1.87 (-0.74)    & 11.41 (0.98)    & -2.29 (-0.51)    \\[3pt]

Anxiety (Dass21) & Demographic & -0.28 (-1.36)    & -0.42 (-5.44***) & 2.38 (1.05)     & 0.88 (1.18)      \\
Age & Demographic & -0.12 (-0.77)    & -0.26 (-2.26**)  & 1.11 (1.09)     & 0.32 (0.53)      \\
Negative affect (PANAS) & Demographic & -0.24 (-1.75*)   & -0.58 (-4.06***) & 1.46 (0.62)     & 0.43 (0.35)      \\
Positive affect (PANAS) & Demographic & -0.01 (-0.17)    & 0.01 (0.06)      & -0.74 (-0.63)   & 0.02 (0.04)      \\
Stress (Dass21) & Demographic & 0.13 (1.11)      & 0.43 (2.30**)    & -1.52 (-0.68)   & -0.67 (-0.49)    \\
Femenine & Demographic & -1.42 (-11.46***) & -0.98 (-5.69***) & -0.53 (-3.70***) & -0.96 (-2.35**) \\
Frequency & Demographic & 0.22 (2.49**)    & 0.31 (2.79***)   & -1.77 (-1.12)   & -0.66 (-1.16)    \\
Knows & Demographic & -0.53 (-1.39)    & -1.03 (-4.11***) & 0.97 (0.21)     & 0.55 (0.28)      \\[3pt]

Distance to junction & Distance to Junction & -21.70 (-1.53)   & -7.19 (-1.59)    & -10.34 (-1.69*) & -4.12 (-1.71*)   \\
Distance to junction x Knows  & Distance to Junction & -0.36 (-1.66*)   & -0.43 (-1.81*)   & -5.50 (-0.91)   & -1.98 (-1.01)    \\[3pt]

Bikeway & Infrastructure & -0.25 (-0.86)    & -0.29 (-0.74)    & -17.54 (-1.21)  & -0.72 (-0.22)    \\[3pt]

CO2 & ContextINO & -0.08 (-0.95)    & -0.10 (-1.01)    & -0.21 (-0.07)   & -0.05 (-0.05)    \\
Humidity & ContextINO & -0.11 (-1.15)    & -0.12 (-1.07)    & 0.31 (0.11)     & 0.09 (0.09)      \\
Temperature & ContextINO & 0.03 (0.29)      & 0.04 (0.25)      & -0.29 (-0.09)   & -0.06 (-0.05)    \\[3pt]

High cyclists activity & GPT &                    & 0.43 (1.39)     &                 & 4.59 (1.48)      \\
High pedestrians activity & GPT &                    & 0.23 (1.15)     &                 & 2.08 (1.13)      \\
High vehicular activity & GPT &                    & -3.41 (-0.96)   &                 & -2.23 (-0.79)    \\
Bad infrastructure & GPT &                    & 16.02 (1.80*)   &                 & 8.68 (1.66*)     \\
Cloudy & GPT &                    & -0.37 (-2.24**) &                 & -1.91 (-1.12)    \\[3pt]

Spectral centroid (audio) & Audio Features &                    &                 & -0.22 (-0.06)   & -0.37 (-0.28)    \\
Spectral contrast (audio) & Audio Features &                    &                 & 17.11 (1.92*)   & 6.04 (2.12**)    \\
RMS (audio) & Audio Features &                    &                 & -9.52 (-1.67*)  & -3.81 (-2.21**)  \\
ZCR (audio) & Audio Features &                    &                 & 0.19 (0.05)     & 0.17 (0.12)      \\

\\[-4pt]
\textbf{Fatigue} \\[3pt]

Constant & Constant & 0.57 (10.92***)  & 0.26 (3.43***)   & 0.41 (1.64)     & 0.56 (8.55***)   \\[3pt]

Anxiety (Dass21) & Demographic & 0.24 (2.47**)    & -0.08 (-2.17**)  & -0.00 (-0.03)   & 0.11 (1.23)      \\
Stress (Dass21) & Demographic & -0.02 (-0.29)    & -0.09 (-1.88*)   & -0.42 (-5.94***) &                  \\
Femenine & Demographic & -1.42 (-11.46***) & -0.98 (-5.69***) & -0.53 (-3.70***) & -0.96 (-2.35**)  \\
BMI & Demographic & 0.02 (0.51)      & 0.23 (4.96***)   & -0.13 (-1.17)   & -0.09 (-0.20)    \\[3pt]

Traveled distance & Distance Traveled & -2.97 (-5.17***) & -2.62 (-6.07***) & -2.58 (-4.98***) & -2.57 (-3.66***) \\
Traveled distance squared & Distance Traveled & 0.55 (1.26)       & 0.54 (1.49)      & 0.55 (1.42)      & 0.56 (1.42)      \\[3pt]

Time & Elapsed Time & 4.74 (25.50***)   & 4.08 (31.33***) & 4.12 (20.16***)  & 4.10 (9.73***)   \\
Time squared & Elapsed Time & -1.07 (-2.24**)   & -0.93 (-2.35**) & -0.94 (-2.31**)  & -0.94 (-2.22**) \\[3pt]

Positive slope & Infrastructure & -0.30 (-2.26**)  & -0.26 (-2.51**)  & -0.25 (-2.12**) & -0.24 (-2.60***) \\[3pt]


Temperature x Humidity & ContextINO & 0.03 (0.34)       & 0.01 (0.22)      & 0.01 (0.10)      & 0.01 (0.09)      \\[3pt]


\hline
\multicolumn{6}{r}{\normalsize\textbf{Note:} \normalsize{Estimated value (Robust t-test). * 90\%, ** 95\%, *** 99\%.}} \\
\hline
\end{tabular}%
}
\end{table}

\section{Glossary of acronyms, subindexes, functions, variables and parameters}
\label{appendix:variables}

In order of introduction:

\begin{table}[]
\begin{tabular}{|l|l|l|}
\hline
\textbf{Term}                              & \textbf{Meaning}                                                                & \textbf{Type}      \\ \hline
VR      & Virtual Reality                                                        & Acronym   \\ \hline
ED      & Electrodermal Activity                                                 & Acronym   \\ \hline
GPS     & Global Position System                                                 & Acronym   \\ \hline
POF     & Points Of Interest                                                     & Acronym   \\ \hline
LLM     & Large Language Models                                                  & Acronym   \\ \hline
ICLV    & Integrated Choice and Latent Variable                                  & Acronym   \\ \hline
DCM     & Discrete Choice Model                                                  & Acronym   \\ \hline
AI      & Artificial Intelligence                                                & Acronym   \\ \hline
RUM     & Random Utility Maximization                                            & Acronym   \\ \hline
$n$                               & Subindex for an individual                                             & Subindex  \\ \hline
$i$                               & Subindex for an alternative                                            & Subindex  \\ \hline
$U_{ni}(\cdot)$                   & Utility function of individual $n$ over alternative $i$                & Function  \\ \hline
$V_{ni}(\cdot)$                   & Systematic utility function of individual $n$ over alternative $i$     & Function  \\ \hline
$\varepsilon_{ni}$                & Random error of individual $n$'s utility function over alternative $i$ & Variable  \\ \hline
$x_{nik}$                         & Individual $n$'s value for attribute $k$ of alternative $i$            & Variable  \\ \hline
$k$                               & Subindex for an attribute                                              & Subindex  \\ \hline
$\beta_k$                         & Taste parameter of attribute $k$                                       & Parameter \\ \hline
$y_{ni}(\cdot)$                   & Indicator function for individual $n$'s choice over alternative $i$    & Function  \\ \hline
$C$                               & Set of alternatives                                                    & Set       \\ \hline
$l$                               & Subindex for a latent value                                            & Subindex  \\ \hline
$x_{nil}$                         & Individual $n$'s value for latent value $l$ of alternative $i$         & Variable  \\ \hline
$\beta_l$                         & Taste parameter of latent value $l$                                    & Parameter \\ \hline
$m$                               & Subindex for latent values indicators                                  & Subindex  \\ \hline
$I_{mni}$                         & Indicator $m$ for latent values of individual $n$ over alternative $i$ & Parameter \\ \hline
ML      & Machine learning                                                       & Acronym   \\ \hline
$f_i(\cdot)$                      & Knowledge-driven term regarding alternative $i$                        & Function  \\ \hline
$\beta$                           & Parameter array for knowledge-driven function                          & Function  \\ \hline
$r_i(\cdot)$                      & Data-driven term regarding alternative $i$                             & Function  \\ \hline
$D_n$                             & Individual $n$'s knowledge-driven variable                             & Variable  \\ \hline
$\omega$                          & Parameters array for data-driven function                              & Parameter \\ \hline
chatGPT & Generative pre-trained transformer chat                                & Acronym   \\ \hline
DASS-21 & Depression Anxiety Stress Scale                                        & Acronym   \\ \hline
PANAS   & Positive and Negative Affect                                           & Acronym   \\ \hline
PPG     & Photoplethysmography                                                   & Acronym   \\ \hline
SKT     & Skin temperature                                                       & Acronym   \\ \hline
HR    &  Heart rate                                                           & Acronym    \\ \hline
SDNN  &  Standard deviation of normal-to-normal                              & Acronym    \\ \hline
RMSSD &  Root mean square of successive differences                        & Acronym    \\ \hline
ECG  & Electrocardiogram                           & Acronym \\ \hline
SCR  & Skin conductance response                   & Acronym \\ \hline
LVD  & Large language model-based video descriptor & Acronym \\ \hline
$t$                       & Subindex for a time instant                                                    & Subindex  \\ \hline
$F_{n,t}(\cdot)$                 & Latent fatigue function for individual $n$ at time instant $t$                 & Function  \\ \hline
$\beta_0^{fat}$           & Constant of latent fatigue function                                            & Parameter \\ \hline
$\beta_{tt}$              & Factor of elapsed time in latent functions                                     & Parameter \\ \hline
$TT_{n,t}$                & Elapsed travel time of individual $n$ at time instant $t$                      & Variable  \\ \hline
$\beta_{dr}$              & Factor of traveled distance in latent functions                                & Parameter \\ \hline
$D_{n,t}$           & Distance traveled by individual $n$ at time instant $t$                        & Variable  \\ \hline                       
\end{tabular}
\end{table}

\begin{table}[]
\begin{tabular}{|l|l|l|}
\hline
\textbf{Term}                      & \textbf{Meaning}                                                                        & \textbf{Type}      \\ \hline
$\beta_{SL}$              & Factor of slope in latent functions                                            & Parameter \\ \hline
$SL_{n,t}$                & Slope of individual $n$ at time instant $t$                                    & Variable  \\ \hline
$\beta_{DF}$              & Factor of demographic features in latent functions                             & Parameter \\ \hline
$DF_{n,t}$                & Demographic features of individual $n$ at time instant $t$                     & Variable  \\ \hline
$\beta_{C}$               & Factor of ContextINO variables in latent functions                             & Parameter \\ \hline
$C_{n,t}$                 & ContextINO variables faced by individual $n$ at time instant $t$ & Vector of \\
                          & (air temperature, relative humidity, ambient noise level, CO2 concentration, & variables\\
                          & and atmospheric pressure) &
                       \\ \hline
$\varepsilon_{n,t}^{fat}$ & Random error for individual $n$ at time instant $t$ in latent fatigue function & Variable  \\ \hline
$A_{n,t}(\cdot)$                 & Latent arousal function for individual $n$ at time isntant $t$                 & Function  \\ \hline
$\beta_0^{act}$           & Constant of latent arousal function                                            & Parameter \\ \hline
$\beta_{af}$              & Factor of audio features in latent functions                                   & Parameter \\ \hline
$AF_{n,t}$                & Audio features of individual $n$ at time instant $t$                           & Variable  \\ \hline
$\beta_{IF}$              & Factor of infrastructure features in latent functions                          & Parameter \\ \hline
$IF_{n,t}$                & Infrastructure features of individual $n$ at time instant $t$                  & Variable  \\ \hline
$\beta_{gpt}$             & Factor of GPT environment variables in latent functions                        & Parameter \\ \hline
$GPT_{n,t}$               & GPT environment variables of individual $n$ at time instant $t$                & Variable  \\ \hline
$\beta_{di}$              & Factor of distance to junctions in latent functions                            & Parameter \\ \hline
$DJ_{n,t}$                & Distance to junctions of individual $n$ at time instant $t$                    & Variable  \\ \hline
$\varepsilon_{n,t}^{act}$ & Random error for individual $n$ at time instant $t$ in latent arousal function & Variable  \\ \hline
$IU_{n,t}^{i}(\cdot)$          & Instant utility function of action $i$ for an individual $n$ at an instant $t$               & Function  \\ \hline
$\beta_0^{i}$           & Constant of actions $i$'s Instant Utility Function                                           & Parameter \\ \hline
$\beta_{DJ}^{i}$        & Factor of Distance to Junction variable in $i$'s Instant Utility Function                    & Parameter \\ \hline
$\beta_{IF}^{i}$        & Factor of Infrastructure variable in $i$'s Instant Utility Function                          & Parameter \\ \hline
$\beta_{A}^{i}$         & Factor of Arousal latent variable in $i$'s Instant Utility Function                          & Parameter \\ \hline
$\beta_{F}^{i}$         & Factor of Fatigue latent variable in $i$'s Instant Utility Function                          & Parameter \\ \hline
$\beta_{GPT}^{i}$       & Factor of Environment GPT variable in $i$'s Instant Utility Function                         & Parameter \\ \hline
$SP_{n,t}$              & Speed of individual $n$ at time instant $t$                & Variable  \\ \hline
$\beta_{SP}^{i}$        & Factor of speed variable in $i$'s Instant Utility Function & Parameter \\ \hline
$\varepsilon_{n,t}^{i}$ & Error in the Instant utility function of action $i$ for an individual $n$ at an instant $t$  & Variable  \\ \hline
TC                      & Tonic component of electrodermal activity                                                    & Acronym   \\ \hline
PC                      & Phasic component of electrodermal activity                                                   & Acronym   \\ \hline
HRV                     & Heart rate variability                                                                       & Acronym   \\ \hline
$TC_{n,t}(\cdot)$              & Tonic component function of individual $n$ at time instant $t$                               & Function  \\ \hline
$\gamma_{A,TC}$           & Factor of arousal variable in Tonic component function                                       & Parameter \\ \hline
$\gamma_{F,TC}$         & Factor of fatigue variable in Tonic component function                                       & Parameter \\ \hline
$\varepsilon_{TC}$          & Error in tonic component function                                                                                       & Parameter \\ \hline
$PC_{n,t}(\cdot)$              & Phasic component function of individual $n$ at time instant $t$                              & Function  \\ \hline
$\gamma_{PC}$           & Factor of arousal variable in Tonic component function                                       & Parameter \\ \hline
$\varepsilon_{PC}$          & Error in phasic component function                                                                                        & Parameter \\ \hline
$HR_{n,t}(\cdot)$              & Heart rate function of individual $n$ at time instant $t$                                     & Function  \\ \hline
$\gamma_{HR}$           & Factor of fatigue variable in heart rate function                                             & Parameter \\ \hline
$\varepsilon_{HR}$          & Error in heart rate function                                                                                        & Parameter \\ \hline
$HRV_{n,t}(\cdot)$             & Heart rate function variability of individual $n$ at time instant $t$                         & Function  \\ \hline
$\gamma_{HRV}$          & Factor of fatigue variable in heart rate variability function                                 & Parameter \\ \hline
$\varepsilon_{HRV}$         & Error in heart rate variability function                                                                                        & Parameter \\ \hline
$\sigma_{\varepsilon}$         & Standard deviation of errors $\varepsilon_{TC}$, $\varepsilon_{PC}$, $\varepsilon_{HR}$ and $\varepsilon_{HRV}$                                                                                 & Variable  \\ \hline
                         
\end{tabular}
\end{table}

\begin{table}[]
\begin{tabular}{|l|l|l|}
\hline
\textbf{Term}                       & \textbf{Meaning}                                                                                      & \textbf{Type}      \\ \hline
$y_{n,t,i}(\cdot)$         & Indicator function for individual $n$'s choice over alternative $i$ at time instant $t$      & Function  \\ \hline
$x_{n,t}$                  & Speed of an individual $n$ at a time $t$                                                     & Variable  \\ \hline
$y_{n,t,i}^{cont}$         & Continuous part of $y_{n,t,i}$, for individual $n$'s choice over alternative $i$ at time $t$ & Variable  \\ \hline
$\mu_i$                    & Mean value of  $y_{n,t,i}^{cont}$ for action $i$                                             & Parameter \\ \hline
$\sigma_i$                 & Standard deviation of  $y_{n,t,i}^{cont}$ for action $i$                                     & Parameter \\ \hline
$\delta_{\text{cont}}(i)$ & Indicator of continuous component of action $i$                                              & Parameter \\ \hline
$\eta_med$                 & Measurement error                                                                            & Variable  \\ \hline
MNL                        & Multinomial logit                                                                            & Acronym   \\ \hline
NLHPC                      & National Laboratory for High Performance Computing                                           & Acronym   \\ \hline
LL(final)                  & Final log-likelihood                                                                         & Acronym   \\ \hline
AIC                        & Akaike Information Criterion                                                                 & Acronym   \\ \hline
BIC                        & Bayesian Information Criterion                                                               & Acronym   \\ \hline
\end{tabular}
\end{table}

\end{document}